\documentclass[showkeys]{revtex4}
\usepackage[T2A]{fontenc}      
\usepackage[russian]{babel}    
\usepackage{amsmath}           
\usepackage{amssymb}           
\usepackage[dvips]{graphicx}   
\usepackage{euscript}
\usepackage{epsfig}

\newcommand{\pd}{\partial}

\begin{document}

\title{╨хфєъЎш  ╧єрэърЁх фы  ухюфхчшўхёъшї эр фхЇюЁьшЁютрээ√ї ёЇхЁрї}

\author{─.╬.\,╤шэшЎ√э}

\affiliation{╠юёъютёъшщ уюёєфрЁёЄтхээ√щ єэштхЁёшЄхЄ шь. ╠.┬. ╦юьюэюёютр\\
ьхїрэшъю-ьрЄхьрЄшўхёъшщ Їръєы№ЄхЄ}

\begin{abstract}
╚чєўр■Єё  ухюфхчшўхёъшх эр ушяхЁяютхЁїэюёЄ ї, сышчъшї ъ ёЄрэфрЁЄэющ (n-1)-ьхЁэющ ёЇхЁх
т n-ьхЁэюь хтъышфютюь яЁюёЄЁрэёЄтх. ╤ыхфє  ╧єрэърЁх, ь√ ЁрёёьрЄЁштрхь чрфрўє т Ёрьърї рэрышЄшўхёъющ ьхїрэшъш
ш шёяюы№чєхь ЄхюЁш■ тючьє∙хэшщ ё Ўхы№■ яюыєўшЄ№ Єюяюыюушўхёъє■ ъырёёшЇшърЎш■ ьэюцхёЄтр
тёхї ухюфхчшўхёъшї эр яютхЁїэюёЄш.
─ы  ¤Єюую ь√ шёяюы№чєхь ыєўхтюх яЁхюсЁрчютрэшх, шчтхёЄэюх т шэЄхуЁры№эющ ухюьхЄЁшш,
ш яюыєўрхь ёшёЄхьє юёЁхфэхээ√ї єЁртэхэшщ фтшцхэш  эр яєрёёюэютющ рыухсЁх єуыютюую ьюьхэЄр,
ъюЄюЁр  юърч√трхЄё  урьшы№Єюэютющ.
╥ръшь юсЁрчюь, юёє∙хёЄты хЄё  рёшьяЄюЄшўхёър  ЁхфєъЎш  шёїюфэющ Єюўэющ ёшёЄхь√ шч $2n-2$
єЁртэхэшщ фы  ухюфхчшўхёъшї ъ юёЁхфэхээющ ёшёЄхьх $2n-4$ єЁртэхэшщ эр ьэюуююсЁрчшш ├Ёрёёьрэр $G(2,n)$.
╤ъюсъш ╧єрёёюэр т эютющ ёшёЄхьх юяЁхфхы ■Єё  рыухсЁющ ╦ш уЁєяя√ $SO(n)$.
┬ трцэ√ї ёыєўр ї фтєьхЁэ√ї, р Єръцх Ё фр ЄЁхїьхЁэ√ї ушяхЁяютхЁїэюёЄхщ яюёЄЁюхээр  ЁхфєъЎш 
яючтюы хЄ т√яюыэшЄ№ Єюяюыюушўхёъє■ ъырёёшЇшърЎш■ ухюфхчшўхёъшї.

\end{abstract}

\keywords{├рьшы№Єюэютр ЁхфєъЎш , ухюфхчшўхёъшх, ыєўхтюх яЁхюсЁрчютрэшх, уЁрёёьрэшрэ, яы■ъъхЁют√ ъююЁфрэрЄ√,
рыухсЁ√ ╦ш}

\date{\today}

\maketitle

\section{┬тхфхэшх}

╚чєўхэшх ухюфхчшўхёъшї ышэшщ эр Ёшьрэют√ї ьэюуююсЁрчш ї тюёїюфшЄ ъ ъырёёшўхёъющ ЁрсюЄх ▀ъюсш,
ъюЄюЁ√щ эр°хы Єюўэюх Ёх°хэшх фы  ухюфхчшўхёъшї эр ¤ыышяёюшфх ьхЄюфрьш рэрышЄшўхёъющ ьхїрэшъш, \cite{jacoby}.
─ры№эхщ°хх яЁюфтшцхэшх с√ыю юёє∙хёЄтыхэю ╧єрэърЁх ё яюью∙№■ яЁшьхэхэш  Єюяюыюушўхёъшї
ьхЄюфют,~\cite{poinc} (Єхъє∙хх ёюёЄю эшх яЁюсыхь√ ёь. т \cite{DNF}).
╚чєўхэшх ухюфхчшўхёъшї юёЄртрыюё№ ёЁхфш ёрь√ї ёыюцэ√ї яЁюсыхь ухюьхЄЁшш, яЁштыхърыю чэрўшЄхы№эюх тэшьрэшх
ш ёыєцшыю яЁютхЁъющ тючьюцэюёЄхщ ьэюушї ьхЄюфют ъръ т ьхїрэшъх, Єръ ш т Єюяюыюушш, \cite{DNF}.

┬ эхфртэхх тЁхь  чрфрўш ю ухюфхчшўхёъшї яюыєўшыш ЁрчтшЄшх т ёт чш ё шёёыхфютрэш ьш шчюьюЁЇшчьют
фшэрьшўхёъшї ёшёЄхь. ╥ръ, ┬.┬. ╩ючыют√ь т \cite{Kozlov} єёЄрэютыхэр ёт ч№ чрфрўш ю ухюфхчшўхёъшї эр фтєьхЁэюь ¤ыышяёюшфх
ёю ёыєўрхь ╩ыхс°р фы  єЁртэхэшщ ╩шЁїуюЇр фтшцхэш  ЄтхЁфюую Єхыр т шфхры№эющ цшфъюёЄш.
└эрыюушўэ√щ Ёхчєы№ЄрЄ фы  ЄЁхїьхЁэюую ¤ыышяёюшфр яюыєўхэ └.┬. ┴юЁшёют√ь ш ╚.╤. ╠рьрхт√ь т \cite{BMpaper},
р фы  $n$-ьхЁэюую ¤ыышяёюшфр -- └.╠. ╧хЁхыюьют√ь т \cite{Perelomov}.

┬ ЁрсюЄх \cite{Poin} ╧єрэърЁх яЁхфыюцшы эют√щ яюфїюф, ёютьх∙р■∙шщ юс∙хшчтхёЄэ√щ ьхЄюф юёЁхфэхэш ,
яЁшьхэ хь√щ т рэрышЄшўхёъющ ьхїрэшъх ёю тЁхьхэ ├рєёёр, ё Єюяюыюушўхёъющ ЄхюЁшхщ фшэрьшўхёъшї ёшёЄхь.
╠хЄюф ╧єрэърЁх юёэют√трхЄё  эр эрсы■фхэшш, ўЄю хёыш ьэюуююсЁрчшх $M^k$
сышчъю ъ ьэюуююсЁрчш■ ${\cal M}^k$, фы  ъюЄюЁюую яЁюсыхьр ухюфхчшўхёъшї фюяєёърхЄ Єюўэюх Ёх°хэшх,
Єю шчєўхэшх ухюфхчшўхёъшї эр $M^k$ ьюцхЄ с√Є№ юёє∙хёЄтыхэю яЁш яюью∙ш ьхЄюфют ЄхюЁшш тючьє∙хэшщ
ъырёёшўхёъющ ьхїрэшъш. ┬ ёыєўрх, ъюуфр ${\cal M}^k$ -- ёЄрэфрЁЄэр  (n-1)-ьхЁэр  ёЇхЁр
$$
    x_1^2 \; + \; x_2^2 \; + \; \ldots \; + \; x_n^2 =1,
$$
ьюцхЄ с√Є№ яЁшьхэхэ ьхЄюф юёЁхфэхэш  ш яюыєўхэр ёшёЄхьр єЁртэхэшщ, ьхэ№°р , ўхь шёїюфэр , ш фр■∙р 
ърўхёЄтхээюх, шыш Єюяюыюушўхёъюх, юяшёрэшх ухюфхчшўхёъшї эр $M^k$. ─рээр  яЁюЎхфєЁр ьюцхЄ с√Є№
эрчтрэр {\it рёшьяЄюЄшўхёъющ ЁхфєъЎшхщ} шёїюфэющ ёшёЄхь√ єЁртэхэшщ.

╧Ёюсыхьр ЁхфєъЎшш шчєўрырё№ т эхсхёэющ ьхїрэшъх уыртэ√ь юсЁрчюь т ёт чш ё чрфрўхщ ЄЁхї Єхы, \cite{whittaker}.
┬ 20-ь тхъх шэЄхЁхё ъ эхщ ёэютр тючЁюё т ёт чш ё ъшЁры№эющ ЄхюЁшхщ яюы .
┬рцэю, ўЄю ъшЁры№эр  ьюфхы№ юс√ўэю юёэютрэр эр уЁєяях ╦ш ёшььхЄЁшш ${\cal G}$, ъюЄюЁр  ёюёЄрты хЄ ёЄЁєъЄєЁэє■
юёэютє хх єЁртэхэшщ ${\cal E}$. ▌Єю юсёЄю Єхы№ёЄтю с√ыю шёяюы№чютрэю ╧юыьрщхЁюь,
ъюЄюЁ√щ єърчры ьхЄюф яюёЄЁюхэш  ьхэ№°хщ ёшёЄхь√ ${\cal R}$ єЁртэхэшщ, ёыхфє■∙шї шч ${\cal E}$
ш юяшё√тр■∙шї фтшцхэшх фшэрьшўхёъшї яхЁхьхээ√ї,  ты ■∙шїё  шэтрЁшрэЄрьш уЁєяя√ ёшььхЄЁшш
${\cal G}$, \cite{pohlmeyer}. ╧юыьрщхЁ т√яюыэшы ЁхфєъЎш■ ъшЁры№эющ ьюфхыш n-яюы ,
т ъюЄюЁющ яюыхтр  яхЁхьхээр  яЁшэшьрхЄ чэрўхэш  эр фтєьхЁэющ ёЇхЁх, ъ єЁртэхэш■ ёшэєё-├юЁфюэ.
╥ръшь юсЁрчюь, ьхЄюф ╧юыьрщхЁр ЄЁхсєхЄ сюурЄющ ёшььхЄЁшш чрфрўш. ┬ ¤Єюь юЄэю°хэшш °шЁюъюх яюых фы 
яЁшьхэхэш  ЁхфєъЎшш ╧юыьрщхЁр яЁхфёЄрты ■Є єЁртэхэш  ╦хуухЄЄр ёяшэютющ фшэрьшъш т ётхЁїЄхъєўхь $^3 He$,
шьх■∙шх уЁєяяє ёшььхЄЁшш $SO(3) \times SU(2) \times U(1)$ (яюфЁюсэюёЄш ёь. т \cite{Leggett}).
╨хфєъЎш  єЁртэхэшщ ╦хуухЄЄр юёюсхээю шэЄхЁхёэр Єхь, ўЄю юэр яЁюыштрхЄ ётхЄ эр Ёюы№, шуЁрхьє■
урьшы№Єюэютющ ёЄЁєъЄєЁющ.
╘ръЄшўхёъш ╧юыьрщхЁ яюърчры, ўЄю ЁхфєъЎш  Єхёэю ёт чрэр ё рыухсЁющ ╧єрёёюэр чрфрўш.
╬фэръю ўрёЄю  тэр  ЇюЁьр ¤Єющ ёт чш эхюўхтшфэр. ─ы  єЁртэхэшщ ╦хуухЄЄр ЁхфєъЎш  ётюфшЄё  ъ эрїюцфхэш■
яюфрыухсЁ√ яєрёёюэютющ рыухсЁ√, юсЁрчютрээющ ёъюсърьш ╧єрёёюэр шёїюфэ√ї фшэрьшўхёъшї яхЁхьхээ√ї
-- ёяшэр ш ярЁрьхЄЁр яюЁ фър (ёь. \cite{golo}).
┬ ¤Єющ ёт чш шфхш ╧єрэърЁх ърёрЄхы№эю ЄхюЁшш тючьє∙хэшщ ш хх Ёюыш т яЁюсыхьх ЁхфєъЎшш юърч√тр■Єё 
тхё№ьр яыюфюЄтюЁэ√ьш. ╧єрэърЁх ЁрёёьрЄЁштры фтєьхЁэє■ яютхЁїэюёЄ№, яюыєўхээє■ шч ёЄрэфрЁЄэющ фтєьхЁэющ ёЇхЁ√
ьрыющ фхЇюЁьрЎшхщ, ш эр°хы єёыютш , яЁш ъюЄюЁ√ї ёє∙хёЄтє■Є {\it чрьъэєЄ√х ухюфхчшўхёъшх}.
┬ эрёЄю ∙хщ ЁрсюЄх, ёыхфє  шфх ь \cite{Poin} ш \cite{pohlmeyer},
ь√ шчєўрхь ухюфхчшўхёъшх эр фхЇюЁьшЁютрээ√ї ёЇхЁрї, Є.х. яютхЁїэюёЄ ї, ьрыю юЄышўр■∙шїё  юЄ ёЄрэфрЁЄэющ ёЇхЁ√,
яюыєўрхь ЁхфєЎшЁютрээ√х єЁртэхэш  ш т√тюфшь Єюяюыюушўхёъшх ёыхфёЄтш  ю ёЄЁєъЄєЁх {\it тёхую}
ьэюцхёЄтр ухюфхчшўхёъшї, эх юс чрЄхы№эю {\it чрьъэєЄ√ї}.

\section{╙Ёртэхэш  ╦руЁрэцр яхЁтюую Ёюфр фы  ухюфхчшўхёъшї}

╠√ ЁрёёьрЄЁштрхь $(n-1)$-ьхЁэ√х ушяхЁяютхЁїэюёЄш т $n$-ьхЁэюь хтъышфютюь яЁюёЄЁрэёЄтх,
сышчъшх ъ ёЄрэфрЁЄэющ ёЇхЁх ш чрфртрхь√х єЁртэхэшхь тшфр:
\begin{equation}
    \label{f:hypersurface}
    \varphi \; \equiv \; x_1^2 \; + \; x_2^2 \; + \; \ldots \; + \; x_n^2 \; -1 \; + \; \varepsilon \; \psi(x_1, x_2, \ldots , x_n) = 0
\end{equation}
уфх $\varepsilon$ -- ьры√щ ярЁрьхЄЁ ш $\psi(x_1, x_2, \ldots, x_n)$ -- ЇєэъЎш  ъююЁфшэрЄ $x_1, x_2, \ldots, x_n$.
╧Ёюсыхьр ЁрчЁх°хэш  фрээющ ёт чш, тююс∙х уютюЁ , тхё№ьр ёыюцэр, ш эрїюцфхэшх ярЁрьхЄЁшчрЎшш ¤Єющ
ушяхЁяютхЁїэюёЄш т юс∙хь ёыєўрх яЁхфёЄрты хЄ сюы№°шх ЄЁєфэюёЄш.
┴юыхх Ёхчєы№ЄрЄштэ√щ яюфїюф юяшЁрхЄё  эр Єю, ўЄю чрфрўр ¤ътштрыхэЄэр ётюсюфэюьє фтшцхэш■ ўрёЄшЎ√,
эр ъюЄюЁюх эрыюцхэр яЁштхфхээр  т√°х ёт ч№, \cite{whittaker}. ╧ю¤Єюьє ь√ ьюцхь чряшёрЄ№
єЁртэхэш  ухюфхчшўхёъшї т ЇюЁьх єЁртэхэшщ ╦руЁрэцр яхЁтюую Ёюфр:
  \begin{equation}
    \label{LagrFirst}
    \ddot{\vec x}
    = \lambda \,
    \frac{\partial \varphi}
     {\partial \vec x}   \mbox{.}
  \end{equation}
╠эюцшЄхы№ ╦руЁрэцр $ \lambda $ чрфрхЄё  ЇюЁьєыющ:
  \begin{equation}
        \label{lamb}
        \lambda = - \frac{\dot{\vec x} \cdot
          \displaystyle
          \frac{\partial^2 \varphi}{\partial
                \vec x^2} \cdot
           \, \dot{\vec x}
          }
         {\left( \displaystyle
         \frac{\partial \varphi}{\partial \vec x}
          \right)^2}
        = - \frac{\dot{\vec x}^2
        + \varepsilon \, \dot{\vec x} \cdot \displaystyle \frac{\partial^2 \psi}{\partial \vec x^2} \cdot \, \dot{\vec x}}
        {\left( \vec x + \varepsilon \displaystyle \frac{\partial \psi}{\partial \vec x} \right)^2},
        \qquad
            \left(\frac{\partial^2 \psi}{\partial \vec x^2}\right)_{ij}
            = \frac{\partial^2 \psi}{\partial x_i \partial x_j}.
\end{equation}
╧Ёш $\varepsilon = 0$  ЄЁрхъЄюЁшш ёшёЄхь√  (\ref{LagrFirst}) -- ухюфхчшўхёъшх эр ёЇхЁх, Є.х. сюы№°шх ъЁєуш.
╧ю¤Єюьє ь√ ьюцхь юяшёрЄ№ ухюфхчшўхёъшх эр ушяхЁяютхЁїэюёЄш, Є.х. фхЇюЁьшЁютрээющ ёЇхЁх,
ё яюью∙№■ ЄхюЁшш тючьє∙хэшщ. ─ы  ¤Єюую ь√ яЁшьхэ хь ъырёёшўхёъшщ ьхЄюф юёЁхфэхэш , \cite{poinc}.

\section{└ёшьяЄюЄшўхёъюх юяшёрэшх ухюфхчшўхёъшї эр фхЇюЁьшЁютрээ√ї ёЇхЁрї}

═р°ш Ёрёёєцфхэш  юёэютрэ√ эр ёыхфє■∙хь эрсы■фхэшш. ╨рёёьюЄЁшь фхЇюЁьшЁютрээє■ ёЇхЁє ${\cal D}$,
юяЁхфхы хьє■ єЁтрэхэшхь (\ref{f:hypersurface}), ш ёЄрэфрЁЄэє■ ёЇхЁє ${\cal S}$, ёююЄтхЄёЄтє■∙є■ ёыєўр■
$\varepsilon = 0$. ╧юёъюы№ъє $\varepsilon \ll 1$, Єю ухюфхчшўхёър 
${\cal G}$ эр ${\cal D}$ ьюцхЄ ЁрёёьрЄЁштрЄ№ё  ъръ ёяшЁры№, тшЄъш ъюЄюЁющ сышчъш ъ яюфїюф ∙шь сюы№°шь
ъЁєурь ${\cal S}$. ╥ръшь юсЁрчюь, ь√ ьюцхь юяшёрЄ№ Єюяюыюуш■ ${\cal G}$ ё яюью∙№■ фтшцхэш  ёююЄтхЄёЄтє■∙шї
сюы№°шї ъЁєуют ${\cal S}$. ╠√ ьюцхь яхЁхтхёЄш ¤Єю ёююсЁрцхэшх т ъюышўхёЄтхээє■ ЇюЁьє ёыхфє■∙шь юсЁрчюь.
╤ Єюяюыюушўхёъющ Єюўъш чЁхэш  ъюэЇшуєЁрЎшюээюх яЁюёЄЁрэёЄтю ${\cal D}$ шёїюфэющ ёшёЄхь√ -- ¤Єю ёЇхЁр.
╙Ёртэхэш  ухюфхчшўхёъшї юяЁхфхы ■Є фшэрьшўхёъє■ ёшёЄхьє эр ${\cal D}$. ╧юёъюы№ъє $\varepsilon$ ьрыю,
Єю ёє∙хёЄтєхЄ ьры√щ яхЁшюф тЁхьхэш $\tau$, т Єхўхэшх ъюЄюЁюую Єюўър яЁюсхурхЄ юЄЁхчюъ ётюхщ ЄЁрхъЄюЁшш,
ёююЄтхЄёЄтє■∙шщ юфэюьє тшЄъє ухюфхчшўхёъющ ёяшЁрыш. ╥ръшь юсЁрчюь, ъюэЇюЁьрЎш  ухюфхчшўхёъющ ёяшЁрыш ьюцхЄ
с√Є№ юяшёрэр яюёЁхфёЄтюь юяшёрэш  ¤тюы■Ўшш сюы№°шї ъЁєуют эр ${\cal S}$, ёююЄтхЄёЄтє■∙шї тшЄърь ёяшЁрыш. \
═ю сюы№°шх ъЁєуш ${\cal S}$ -- ¤Єю ёхўхэш  ${\cal S}$ фтєьхЁэ√ьш яыюёъюёЄ ьш, яЁюїюф ∙шьш ўхЁхч ЎхэЄЁ ${\cal S}$.
╧ю¤Єюьє ЄЁрхъЄюЁш , ёююЄтхЄёЄтє■∙р  ухюфхчшўхёъющ эр ${\cal S}$, яюЁюцфрхЄ ЄЁрхъЄюЁш■ эр ьэюуююсЁрчшш
тёхї фтєьхЁэ√ї яыюёъюёЄхщ т $n$-ьхЁэюь хтъышфютюь яЁюёЄЁрэёЄтх. ▌Єю ьэюуююсЁрчшх, эрч√трхьюх
ьэюуююсЁрчшхь ├Ёрёёьрэр $G(2, n)$, шчтхёЄэю т ухюьхЄЁшш ё ёхЁхфшэ√ XIX тхър. ╘ръЄшўхёъш эр° яюфїюф
ёыхфєхЄ шфх ь ╘. ╩ыхщэр ш ▐. ╧ы■ъъхЁр, \cite{klein}, \cite{Gelf},
ъюЄюЁ√х яЁхфыюцшыш ьхЄюф ъюэёЄЁєшЁютрэш  эют√ї ухюьхЄЁшўхёъшї юс·хъЄют шч яюфьэюуююсЁрчшщ фрээюую ьэюуююсЁрчш .
╤ Єюяюыюушўхёъющ Єюўъш чЁхэш , ь√ ЁрёёьрЄЁштрхь яхЁхїюф юЄ ёшёЄхь√, юяЁхфхыхээющ эр Єюяюыюушўхёъющ
$(n-1)$-ьхЁэющ ёЇхЁх ъ ёшёЄхьх эр ьэюуююсЁрчшш ├Ёрёёьрэр $G(2,n)$. ╧хЁхїюф шыы■ёЄЁшЁєхЄё  ёыхфє■∙хщ фшруЁрььющ:
$$
   \mbox{ухюфхчшўхёъшх эр фхЇюЁьшЁютрээющ ёЇхЁх}  \Longrightarrow \mbox{фшэрьшўхёър  ёшёЄхьр}\ {\cal H}\
\mbox{эр ьэюуююсЁрчшш ├Ёрёёьрэр}
$$

─ы  юёє∙хёЄтыхэш  шчыюцхээющ ёїхь√ эхюсїюфшью ъюышўхёЄтхээю юяшёрЄ№ фтшцхэшх Єхъє∙хую сюы№°юую ъЁєур,
яЁшсышцр■∙хую фрээ√щ тшЄюъ ухюфхчшўхёъющ ёяшЁрыш. ╧Ёхцфх тёхую, ЄЁхсєхЄё  т√сЁрЄ№ ёяюёюс
чрфрэш  яюыюцхэш  ¤Єюую ъЁєур ўшёырьш -- хую ъююЁфшэрЄрьш. ▌Єю ючэрўрхЄ ттхфхэшх ъююЁфшэрЄ эр ьэюуююсЁрчшш
├Ёрёёьрэр.
╧Ёш ¤Єюь ёє∙хёЄтхээю, ўЄю яюыюцхэшх Єхъє∙хую сюы№°юую ъЁєур юяЁхфхы хЄё  ёюёЄю эшхь ўрёЄшЎ√, Є.х. хх Ёрфшєё-тхъЄюЁюь $\vec x$ ш ёъюЁюёЄ№■ $\dot{\vec x}$, р шьхээю Єхъє∙шщ сюы№°ющ ъЁєу -- ¤Єю ёхўхэшх ёЇхЁ√
фтєьхЁэющ яыюёъюёЄ№■, яЁюїюф ∙хщ ўхЁхч ЎхэЄЁ ёЇхЁ√ ш ёюфхЁцр∙хщ ¤Єш фтр тхъЄюЁр.
╨рёёьюЄЁшь т ърўхёЄтх ъююЁфшэрЄ ¤Єющ яыюёъюёЄш (р чэрўшЄ ш сюы№°юую ъЁєур) ёыхфє■∙шщ эрсюЁ ўшёхы:
\begin{equation}
l_{ij} = x_i \,\dot x_j - x_j \,\dot x_i, \ i,j = 1 \ldots n.
\label{MomDef}
\end{equation}
▌Єю ъюьяюэхэЄ√ ьрЄЁшЎ√ $n$-ьхЁэюую єуыютюую ьюьхэЄр,  ты ■∙хуюё  юсюс∙хэшхь єуыютюую ьюьхэЄр
т ЄЁхїьхЁэюь хтъышфютюь яЁюёЄЁрэёЄтх. ╧юёыхфэшщ ўр∙х чряшё√тр■Є т тшфх тхъЄюЁр $\vec L$, 
ёт чрээюую ё ьрЄЁшЎхщ (\ref{MomDef}) ёююЄэю°хэшхь $L_i = \varepsilon_{ijk} l_{jk}$.
╬ўхтшфэю, $l_{ji} = - l_{ij}$, Є.х. $l_{ij}$ юсЁрчє■Є ъюёюёшььхЄЁшўхёъє■ ьрЄЁшЎє.
┬рцэ√щ ЇръЄ ёюёЄюшЄ т Єюь, ўЄю ттхфхээ√х тхышўшэ√ (\ref{MomDef}) ёютярфр■Є ё Єръ эрч√трхь√ьш
{\it яы■ъъхЁют√ьш ъююЁфшэрЄрьш} фрээющ фтєьхЁэющ яыюёъюёЄш.
╘ръЄшўхёъш, яю ЇюЁьєырь (\ref{MomDef}) ь√ т√ўшёы хь ¤Єш ъююЁфшэрЄ√ фы  Єхъє∙хую сюы№°юую ъЁєур
шёїюф  шч ¤ыхьхэЄют ЄЁрхъЄюЁшш ўрёЄшЎ√. ┬ ёыєўрх, ъюуфр ъююЁфшэрЄ√ яюёЄю ээ√ $l_{ij} = const$,
сюы№°ющ ъЁєу яюъюшЄё , ш ЄЁрхъЄюЁш  тёх тЁхь  юёЄрхЄё  тсышчш юфэющ ш Єющ цх яыюёъюёЄш.
╬ърч√трхЄё , ўЄю эх тёх тхышўшэ√ $l_{ij}, \; i<j$ эхчртшёшь√. ╠юцэю яюърчрЄ№, \cite{Gelf},
ўЄю юэш єфютыхЄтюЁ ■Є ёыхфє■∙шь {\it ёююЄэю°хэш ь ╧ы■ъъхЁр}:
\begin{equation}
l_{j [k_1} l_{k_2 k_3]} \equiv
\frac13 (l_{j k_1} l_{k_2 k_3} - l_{j k_2} l_{k_1 k_3} + l_{j k_3} l_{k_1 k_2}) = 0.
\label{Rels}
\end{equation}
╩Ёюьх Єюую, ъююЁфшэрЄ√ (\ref{MomDef}) юфэюЁюфэ√, Є.х. єьэюцхэшх тёхї ъююЁфшэрЄ эр
юфэю ўшёыю яЁштюфшЄ ъ Єющ цх яыюёъюёЄш. ╧ю¤Єюьє ьюцэю юуЁрэшўшЄ№ё  ьрЄЁшЎрьш
ё хфшэшўэющ ёєььющ ътрфЁрЄют ¤ыхьхэЄют: 
\begin{equation}
\sum_{i,j}l_{ij}^2 = 1.
\label{Unit}
\end{equation}
╬ЄюцфхёЄты   яюёых ¤Єюую ьрЄЁшЎ√ $l_{ij}$ ш $-l_{ij}$, яюыєўрхь
ьэюуююсЁрчшх ├Ёрёёьрэр $G(2,n)$ ЁрчьхЁэюёЄш $2 n - 4$, \cite{Gelf}.

╧юёъюы№ъє фхЇюЁьшЁютрээр  ёЇхЁр сышчър ъ ёЄрэфрЁЄэющ, Єю яыюёъюёЄ№, юяЁхфхы ■∙р  Єхъє∙шщ сюы№°ющ ъЁєу, фтшцхЄё  ьхфыхээю, ш хх ъююЁфшэрЄ√ (\ref{MomDef}), Є.х. ъюьяюэхэЄ√ єуыютюую ьюьхэЄр ўрёЄшЎ√,  ты ■Єё  ьхфыхээ√ьш
яхЁхьхээ√ьш чрфрўш т Єюь ёь√ёых, ўЄю яЁш юсїюфх юфэюую тшЄър ЄЁрхъЄюЁшш шї шчьхэхэшх ьрыю.
─рыхх ь√ юяшё√трхь ¤Єє ьхфыхээє■ фшэрьшъє, шёяюы№чє  ьхЄюф юёЁхфэхэш .

\section{╙Ёртэхэш  фы  єуыютюую ьюьхэЄр}

╚ёяюы№чє  єЁртэхэш  ╦руЁрэцр яхЁтюую Ёюфр (\ref{LagrFirst}), яюыєўрхь ёыхфє■∙шх єЁртэхэш  фы  ъюьяюэхэЄ
єуыютюую ьюьхэЄр:
\begin{equation}
\dot{l}_{ij} = \varepsilon \, \lambda \left(x_i \frac{\pd}{\pd x_j} - x_j \frac{\pd}{\pd x_i} \right) \psi(\vec x),
\label{mom0}
\end{equation}
уфх $\psi(\vec x)$ -- ЇєэъЎш , юяЁхфхы ■∙р  фхЇюЁьрЎш■ ёЇхЁ√, (\ref{f:hypersurface}).
╧Ёш $\varepsilon = 0$ шьххь $\dot{l}_{ij} = 0$, ўЄю ёююЄтхЄёЄтєхЄ ёыєўр■ ёЄрэфрЁЄэющ ёЇхЁ√, ъюуфр
сюы№°ющ ъЁєу яюъюшЄё .

╟рьхЄшь, ўЄю яЁртр  ўрёЄ№ єЁртэхэшщ (\ref{mom0}) яюыєўрхЄё  шч ЇєэъЎшш фхЇюЁьрЎшш ёЇхЁ√ $\psi(\vec x)$
ё яюью∙№■ яЁшьхэхэш  ёыхфє■∙хую юяхЁрЄюЁр:
\begin{equation}
\displaystyle m_{ij} = x_i \frac{\pd}{\pd x_j} - x_j \frac{\pd}{\pd x_i}.
\label{Qmom}
\end{equation}
─рээ√щ юяхЁрЄюЁ шчтхёЄхэ ъръ ъюьяюэхэЄр юяхЁрЄюЁр єуыютюую ьюьхэЄр т ътрэЄютющ ьхїрэшъх.
┬рцэ√ь хую ётющёЄтюь, шёяюы№чєхь√ь т фры№эхщ°хь,  ты хЄё  ёыхфє■∙хх яЁхфёЄртыхэшх
 т тшфх юяхЁрЄюЁр схёъюэхўэю ьрыюую яютюЁюЄр.
╨рёёьюЄЁшь тЁр∙хэшх т яыюёъюёЄш $x_i x_j$ эр єуюы $\phi$.
╬эю фхщёЄтєхЄ эр тхъЄюЁ√ ъръ єьэюцхэшх эр шчтхёЄэє■ юЁЄюуюэры№эє■ ьрЄЁшЎє яютюЁюЄр $R_{ij}(\phi)$.
─хщёЄтшх ¤Єюую тЁр∙хэш  эр ЇєэъЎшш $f(\vec x)$ юяЁхфхы хЄё  ъръ:
$$
R_{ij}(\phi) f(\vec x) = f\left(R^{-1}_{ij}(\phi) \vec x\right).
$$
╥юуфр юяхЁрЄюЁ $m_{ij}$ -- ухэхЁрЄюЁ ¤Єюую тЁр∙хэш  ЇєэъЎшщ:
$$
m_{ij} f(\vec x) = - \left. \left( \frac{d}{d\phi} f\left(R^{-1}_{ij}(\phi) \vec x\right) \right) \right|_{\phi = 0}.
$$
┬ єЁртэхэш ї фы  ьюьхэЄр (\ref{mom0}) ь√ єфхЁцштрхь Єюы№ъю ўыхэ√ 1-ую яюЁ фър яю $\varepsilon$,
Є.х. ЁрёёьрЄЁштрхь 1-щ яюЁ фюъ ЄхюЁшш тючьє∙хэшщ.
╩Ёюьх Єюую, фы  ъЁрЄъюёЄш фрыхх ь√ яЁхфяюырурхь хфшэшЎ√ шчьхЁхэш  т√сЁрээ√ьш Єръшь юсЁрчюь, ўЄю уыртэр 
ўрёЄ№ ьэюцшЄхы  ╦руЁрцэр (\ref{lamb}) $\lambda_0 = -1$ (ёююЄтхЄёЄтєхЄ хфшэшўэющ ёъюЁюёЄш $\dot{\vec x}^2 = 1$,
Є.х. эрЄєЁры№эющ ярЁрьхЄЁшчрЎшш ухюфхчшўхёъющ).
╤ єўхЄюь ¤Єюую єЁртэхэш  фы  єуыютюую ьюьхэЄр (\ref{mom0}), ё шёяюы№чютрэшхь ттхфхээюую юсючэрўхэш 
юяхЁрЄюЁр (\ref{Qmom}), чряшё√тр■Єё  т тшфх:
\begin{equation}
\dot{l}_{ij} = - \varepsilon \, m_{ij} \, \psi(\vec x),
\label{mom}
\end{equation}
╚Єръ, ь√ яюыєўшыш єЁртэхэш  (\ref{mom}), юяшё√тр■∙шх фшэрьшъє ъюьяюэхэЄ єуыютюую ьюьхэЄр
(\ref{MomDef}) ўрёЄшЎ√ яЁш хх фтшцхэшш яю ухюфхчшўхёъющ.

\section{╬ёЁхфэхэшх єЁртэхэшщ фы  ьюьхэЄр}

╧юёъюы№ъє яЁрт√х ўрёЄш єЁртэхэшщ (\ref{mom}) фы  ъюьяюэхэЄ єуыютюую ьюьхэЄр $l_{ij}$
яЁюяюЁЎшюэры№э√ ьрыюьє ярЁрьхЄЁє $\varepsilon$,
Єю фтшцхэшх сюы№°юую ъЁєур  ты хЄё  ьхфыхээ√ь яю ёЁртэхэш■ ё фтшцхэшхь ўрёЄшЎ√ тфюы№ тшЄър ухюфхчшўхёъющ.
╚э√ьш ёыютрьш, $l_{ij}$ -- ьхфыхээ√х яхЁхьхээ√х. ─ы  эхтючьє∙хээюую ёыєўр  ёЄрэфрЁЄэющ ёЇхЁ√ юэш  ты ■Єё  шэЄхуЁрырьш фтшцхэш . ╧ю¤Єюьє ь√ ьюцхь яЁшьхэшЄ№ ъырёёшўхёъшщ ьхЄюф юёЁхфэхэш , \cite{poinc}.
╧Ёрт√х ўрёЄш єЁртэхэшщ (\ref{mom}) ёюфхЁцрЄ ьхфыхээю ьхэ ■∙шхё  ъюьяюэхэЄ√, чртшё ∙шх юЄ яюыюцхэш 
Єхъє∙хую сюы№°юую ъЁєур, яЁшсышцр■∙хую фрээ√щ тшЄюъ ухюфхчшўхёъющ, ш с√ёЄЁ√х юёЎшыышЁє■∙шх ўыхэ√,
чртшё ∙шх юЄ яюыюцхэш  ўрёЄшЎ√ эр тшЄъх. ╧ЁшэЎшя юёхфэхэш  єЄтхЁцфрхЄ, ўЄю ъЁєяэюьрё°Єрсэ√х шчьхэхэш 
Ёх°хэш  чртшё Є Єюы№ъю юЄ ьхфыхээ√ї ъюьяюэхэЄ, р с√ёЄЁю юёЎшыышЁє■∙шх ёырурхь√х яЁштюф Є
ъ ьры√ь ъюыхсрэш ь Ёх°хэш  юъюыю ъЁштющ, юяЁхфхы хьющ ьхфыхээющ ўрёЄ№■.
╥ръшь юсЁрчюь, фы  юяЁхфхыхэш  Єюяюыюушш ухюфхчшўхёъшї ь√ ьюцхь ЁрёёьюЄЁхЄ№ єЁртэхэш  фы  ьюьхэЄр
(\ref{mom}) схч юёЎшыышЁє■∙хщ ўрёЄш. ─ы  ¤Єюую эхюсїюфшью юёЁхфэшЄ№ шї яю яхЁшюфє Єюўэюую Ёх°хэш 
эхтючьє∙хээющ ёшёЄхь√ -- ЁртэюьхЁэюую фтшцхэш  яю сюы№°юьє ъЁєує:
\begin{equation}
    \vec x_{\hat l}(t) = \cos{t} \ \vec e_1(\hat l) + \sin{t} \  \vec e_2(\hat l),
\label{exact}
\end{equation}
уфх $\hat l$ -- ¤Єю фрээр  (яюёЄю ээр ) ьрЄЁшЎр єуыютюую ьюьхэЄр,
$\vec e_1(\hat l), \vec e_2(\hat l)$ -- тхъЄюЁ√, юсЁрчє■∙шх юЁЄюэюЁьшЁютрээ√щ срчшё т фтєьхЁэющ яыюёъюёЄш,
т ъюЄюЁющ яЁюшёїюфшЄ фтшцхэшх ё фрээ√ь єуыют√ь ьюьхэЄюь
(ъюэъЁхЄэ√щ т√сюЁ тхъЄюЁют $\vec e_1(\hat l), \vec e_2(\hat l)$ эх тыш хЄ эр Ёхчєы№ЄрЄ юёЁхфэхэш ).

╘юЁьєыр (\ref{exact}) юяЁхфхы хЄ ЁртэюьхЁэюх фтшцхэшх яю сюы№°юьє ъЁєує, ряяЁюъёшьшЁє■∙хьє Єхъє∙шщ
тшЄюъ ухюфхчшўхёъющ ёяшЁрыш.
╤Ёхфэхх чэрўхэшх ЇєэъЎшш $f(\vec x)$ яю яхЁшюфє ¤Єюую Ёх°хэш  т√ЁрцрхЄё  ЇюЁьєыющ:
\begin{equation}
\left\langle f(\vec x) \right\rangle_{\vec x_{\hat l}(t)} =
\frac{1}{2 \pi} \int_0^{2 \pi} f(\vec x_{\hat l}(t)) \,dt
\label{averx}
\end{equation}
╨хчєы№ЄрЄ юёЁхфэхэш  чртшёшЄ Єюы№ъю юЄ ьрЄЁшЎ√ єуыютюую ьюьхэЄр $\hat l$.
╥хь ёрь√ь, юёЁхфэхэшх єЁртэхэшщ (\ref{mom}) фрхЄ чрьъэєЄє■ ёшёЄхьє єЁртэхэшщ фы  ъюьяюэхэЄ
єуыютюую ьюьхэЄр $l_{ij}$:
\begin{equation}
\displaystyle
\dot{l}_{ij} = - \varepsilon \left\langle m_{ij} \psi(\vec x) \right\rangle_{\vec x_{\hat l}},
\label{momAv0}
\end{equation}
уфх $\psi(\vec x)$ -- ЇєэъЎш , юяЁхфхы ■∙р  фхЇюЁьрЎш■ ёЇхЁ√, (\ref{f:hypersurface}),
$m_{ij}$ -- юяхЁрЄюЁ, юяЁхфхыхээ√щ т (\ref{Qmom}), шчтхёЄэ√щ, ъръ юяхЁрЄюЁ єуыютюую ьюьхэЄр т ътрэЄютющ
ьхїрэшъх.
═ряюьэшь, ўЄю тхышўшэ√ $l_{ij}$  ты ■Єё  яы■ъъхЁют√ьш ъююЁфшэрЄрьш фтєьхЁэющ яыюёъюёЄш сюы№°юую ъЁєур,
ряяЁюъёшьшЁє■∙хую Єхъє∙шщ тшЄюъ ухюфхчшўхёъющ ёяшЁрыш.
╧ю¤Єюьє ЇръЄшўхёъш ёшёЄхьр (\ref{momAv0}) чрфрэр эр ьэюуююсЁрчшш ├Ёрёёьрэр $G(2,n)$.

\section{╤т ч№ ё шэЄхуЁры№эющ ухюьхЄЁшхщ}

\begin{figure}
  \begin{center}
    \includegraphics[width = 250bp]{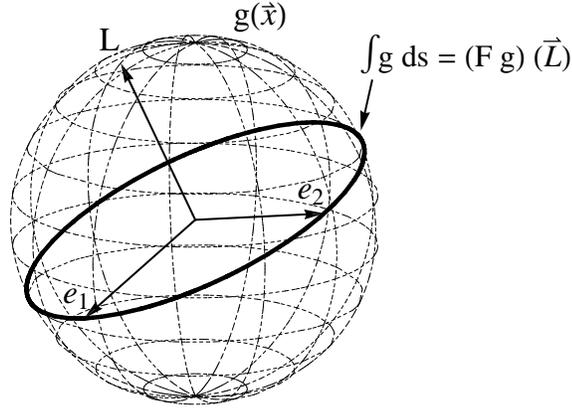}
    \caption{╧ЁхюсЁрчютрэшх ╘єэър-╠шэъютёъюую ЇєэъЎшш $g(\vec x)$ эр хфшэшўэющ ёЇхЁх,
                 тч Єюх т Єюўъх $\vec L$, -- ¤Єю шэЄхуЁры юЄ $g$ яю сюы№°юьє ъЁєує, яхЁяхэфшъєы Ёэюьє ъ $\vec L$.}
    \label{fig1}
  \end{center}
\end{figure}

┬рцэюх эрсы■фхэшх ёюёЄюшЄ т Єюь, ўЄю шёяюы№чютрээр  яЁюЎхфєЁр юёЁхфэхэш  (\ref{averx})
шьххЄ ёт ч№ ё яЁхюсЁрчютрэш ьш, ЁрёёьрЄЁштрхь√ьш т {\it шэЄхуЁры№эющ ухюьхЄЁшш}, \cite{Gelf}, \cite{HelgRad}.
╚ьхээю, юэю ¤ътштрыхэЄэю Єръ эрч√трхьюьє {\it ыєўхтюьє яЁхюсЁрчютрэш■} $J$,
ёюяюёЄрты ■∙хьє ЇєэъЎшш $f(\vec x)$ эр ёЄрэфрЁЄэющ ёЇхЁх хх шэЄхуЁры√ яю тёхтючьюцэ√ь сюы№°шь ъЁєурь.
╘єэъЎш -юсЁрч $J f$ ыєўхтюую яЁхюсЁрчютрэш  юяЁхфхыхэр эр ьэюцхёЄтх сюы№°шї ъЁєуют, шыш, ¤ътштрыхэЄэю, эр ьэюцхёЄтх
ёюфхЁцр∙шї шї фтєьхЁэ√ї яыюёъюёЄхщ, яЁюїюф ∙шї ўхЁхч ЎхэЄЁ ёЇхЁ√, Є.х. эр ьэюуююсЁрчшш ├Ёрёёьрэр $G(2,n)$.
╚Єръ, ыєўхтюх яЁхюсЁрчютрэшх чрфрхЄё  ЇюЁьєыющ:
\begin{equation}
(J f) (\hat p) = \int_0^{2 \pi} f\left(\cos{t} \ \vec e_1(\hat p) + \sin{t} \ \vec e_2(\hat p)\right) \, dt.
\label{Xray}
\end{equation}
 ╟фхё№ $\hat p$ -- ьрЄЁшЎр яы■ъъхЁют√ї ъююЁфшэрЄ фтєьхЁэющ яыюёъюёЄш;
$\vec e_1(\hat p), \vec e_2(\hat p)$ -- юЁЄюэюЁьшЁютрээ√щ срчшё т ¤Єющ яыюёъюёЄш.
╨шё. \ref{fig1} шыы■ёЄЁшЁєхЄ ёыєўрщ $n=3$, Є.х. ёыєўрщ фтєьхЁэющ ёЇхЁ√, т ъюЄюЁюь ыєўхтюх яЁхюсЁрчютрэшх
эюёшЄ эрчтрэшх яЁхюсЁрчютрэш  ╘єэър-╠шэъютёъюую, ёь. \cite{Gelf}.

╚ч ёЁртэхэш  ЇюЁьєы (\ref{Xray}) ш (\ref{exact}), (\ref{averx}) ь√ тшфшь, ўЄю юёЁхфэхэшх (\ref{averx})
т√ЁрцрхЄё  ўхЁхч ыєўхтюх яЁхюсЁрчютрэшх яю ЇюЁьєых:
$$
\left\langle f(\vec x) \right\rangle_{\vec x_{\hat l}(t)} = \frac{1}{2 \pi} (J f) (\hat l) .
$$

╥ръшь юсЁрчюь, юёЁхфэхээр  ёшёЄхьр (\ref{mom}) фы  єуыютюую ьюьхэЄр ьюцхЄ с√Є№ чряшёрэр т тшфх:
\begin{equation}
\displaystyle
\dot{l}_{ij}
= - \frac{\varepsilon}{2 \pi} \,
\left( J \circ m_{ij} \ \psi \right) \, (\hat l),
\label{momAv}
\end{equation}
╟фхё№ $\psi(\vec x)$ -- ЇєэъЎш , юяЁхфхы ■∙р  фхЇюЁьрЎш■ ёЇхЁ√, (\ref{f:hypersurface});
$m_{ij}$ -- юяхЁрЄюЁ, юяЁхфхыхээ√щ т (\ref{Qmom}), шчтхёЄэ√щ, ъръ юяхЁрЄюЁ єуыютюую ьюьхэЄр т ътрэЄютющ
ьхїрэшъх; $J$ -- ыєўхтюх яЁхюсЁрчютрэшх, юяЁхфхыхээюх т (\ref{Xray}).

╧юыєўхээ√х єЁртэхэш  (\ref{momAv}) юяшё√тр■Є юёЁхфэхээє■ фшэрьшъє єуыютюую ьюьхэЄр ўрёЄшЎ√,
фтшцє∙хщё  яю ухюфхчшўхёъющ.
╧Ёш ¤Єюь ъюьяюэхэЄ√ ьюьхэЄр $l_{ij}$  ты ■Єё  яы■ъъхЁют√ьш ъююЁфшэрЄрьш фтєьхЁэющ яыюёъюёЄш 
Єхъє∙хую сюы№°юую ъЁєур,
ш, ёююЄтхЄёЄтхээю, ёшёЄхьр (\ref{momAv}) чрфрэр эр ьэюуююсЁрчшш ├Ёрёёьрэр $G(2,n)$.
╥хь ёрь√ь, юёє∙хёЄтыхэр рёшьяЄюЄшўхёър  ЁхфєъЎш  шёїюфэющ ёшёЄхь√ фы  ухюфхчшўхёъшї
ё Їрчютющ ЁрчьхЁэюёЄ№■ $2n-2$ ъ юёЁхфэхээющ ёшёЄхьх $2n-4$ єЁртэхэшщ фы  ьюьхэЄр.
┬рцэ√щ ЇръЄ ёюёЄюшЄ т Єюь, ўЄю фрээр  ёшёЄхьр шьххЄ урьшы№Єюэютє ёЄЁєъЄєЁє, ъюЄюЁр  юсёєцфрхЄё 
т ёыхфє■∙хь Ёрчфхых.

\section{├рьшы№Єюэютр ёЄЁєъЄєЁр ЁхфєЎшЁютрээющ ёшёЄхь√ фы  єуыютюую ьюьхэЄр}

├рьшы№Єюэютр ЇюЁьєышЁютър юёЁхфэхээ√ї єЁртэхэшщ фы  єуыютюую ьюьхэЄр (\ref{momAv}) ЄЁхсєхЄ чрфрэш 
ёъюсюъ ╧єрёёюэр ьхцфє фшэрьшўхёъшьш яхЁхьхээ√ьш $l_{ij}$ ш єърчрэш  урьшы№Єюэшрэр.

╤ъюсъш ╧єрёёюэр фы  єуыютюую ьюьхэЄр юяЁхфхы ■Єё  рыухсЁющ ╦ш $so(n)$ ш шьх■Є тшф:
\begin{equation}
\{l_{ij}, l_{pq} \} = \delta_{ip}l_{jq} + \delta_{jp}l_{qi} + \delta_{iq}l_{pj} + \delta_{jq}l_{ip}.
\label{momPois}
\end{equation}

═шцх ь√ яюърч√трхь, ўЄю урьшы№Єюэшрэюь ёшёЄхь√ (\ref{momAv}) ёыєцшЄ ёыхфє■∙р  ЇєэъЎш :
\begin{equation}
H(\hat l) = \frac{\varepsilon}{2\pi} J \psi,
\label{momHam}
\end{equation}
ъюЄюЁр  яюыєўрхЄё  яЁшьхэхэшхь ыєўхтюую яЁхюсЁрчютрэш  $J$, (\ref{Xray}), ъ ЇєэъЎшш $\psi(\vec x)$,
юяЁхфхы ■∙хщ фхЇюЁьрЎш■ ёЇхЁ√.

─ы  фюърчрЄхы№ёЄтр єЄтхЁцфхэш  ю урьшы№Єюэютющ ёЄЁєъЄєЁх эрь яюЄЁхсєхЄё  ёыхфє■∙хх ётющёЄтю
ёъюсюъ ╧єрёёюэр (\ref{momPois}).

┬шф ёъюсюъ (\ref{momPois}) ёютярфрхЄ ё ъюььєЄрЄюЁюь т рыухсЁх ╦ш $so(n)$ срчшёэ√ї ъюёюёшььхЄЁшўхёъшї ьрЄЁшЎ
$E_{ij}$, юяЁхфхы хь√ї ЇюЁьєыющ $(E_{ij})_{ab}= - \delta_{ia}\delta_{jb} + \delta_{ib}\delta_{ja}$.
╚ьхээю, шї ъюьєЄрЄюЁ шьххЄ тшф:
\begin{equation}
[E_{ij}, E_{pq}] = \delta_{ip}E_{jq} + \delta_{jp}E_{qi} + \delta_{iq}E_{pj} + \delta_{jq}E_{ip}.
\label{commSO}
\end{equation}
╬э ьюцхЄ с√Є№ т√Ёрцхэ ўхЁхч ёЄЁєъЄєЁэ√х ъюэёЄрэЄ√ $ C_{ij,pq,ab}$ рыухсЁ√  ╦ш $so(n)$:
$$
[E_{ij}, E_{pq}] = C_{ij,pq,ab} E_{ab},
$$
ъюЄюЁ√х шьх■Є тшф:
$$
C_{ij,pq,ab} = \frac12 \left(
\delta _{i p} \left(\delta_{a j} \delta_{b q} - \delta_{a q} \delta_{b j}\right)+
\delta _{j p} \left(\delta_{a q} \delta_{b i} - \delta_{a i} \delta_{b q}\right)+
\delta _{i q} \left(\delta_{a p} \delta_{b j} - \delta_{a j} \delta_{b p}\right)+
\delta _{j q} \left(\delta_{a i} \delta_{b p} - \delta_{a p} \delta_{b i}\right)
\right).
$$
╦хуъю яЁютхЁшЄ№, ўЄю ёЄЁєъЄєЁэ√х ъюэёЄрэЄ√ єфютыхЄтюЁ ■Є ёююЄэю°хэш■ $ C_{ab,pq,ij} = - C_{ij,pq,ab} $,
ъюЄюЁюх сєфхЄ шёяюы№чютрэю т фры№эхщ°хь.

╩Ёюьх Єюую, эрь яюЄЁхсєхЄё  ёфхфє■∙хх {\it юёэютэюх ъюььєЄрЎшюээюх ёююЄэю°хэшх фы  ыєўхтюую яЁхюсЁрчютрэш :}
\begin{equation}
J \circ m_{ij} = c_{ij} \circ J,
\label{comm}
\end{equation}
уфх $J$ -- ыєўхтюх яЁхюсЁрчютрэшх, юяЁхфхыхээюх т (\ref{Xray}), $m_{ij}$ -- юяхЁрЄюЁ,
юяЁхфхыхээ√щ т (\ref{Qmom}), $ c_{ij} $ -- юяхЁрЄюЁ, фхщёЄтє■∙шщ эр ЇєэъЎш■ юЄ єуыютюую ьюьхэЄр
$f(\hat l)$ ъръ тч Єшх хх ёъюсъш ╧єрёёюэр ё $l_{ij}$:
$$
c_{ij} \, f(\hat l) = \left\{ f(\hat l), l_{ij} \right\}.
$$
─ы  т√тюфр ёююЄэю°хэш  (\ref{comm}) шёяюы№чєхь ёыхфє■∙хх ётющёЄтю,
юяЁхфхы ■∙хх фхщёЄтшх тЁр∙хэшщ эр юсЁрч ыєўхтюую яЁхюсЁрчютрэш .
╧юфёЄрты   т юяЁхфхыхэшх ыєўхтюую яЁхюсЁрчютрэш  (\ref{Xray}) тьхёЄю $f(\vec x)$ ЇєэъЎш■, яюыєўхээє■ шч эхх тЁр∙хэшхь $f\left(R^{-1}_{ij}(\phi) \vec x\right)$, яюыєўрхь:
\begin{equation}
\left(J \, f\left(R^{-1}_{ij}(\phi) \vec x\right)\right) (\hat l) =
\int_0^{2 \pi} f\left(R^{-1}_{ij}(\phi) \left(\cos{t} \ \vec e_1(\hat l) + \sin{t} \ \vec e_2(\hat l)\right)\right) \, dt.
\label{XrayRot}
\end{equation}
┼ёыш $C$ -- сюы№°ющ ъЁєу ё яы■ъъхЁют√ьш ъююЁфшэрЄрьш $\hat l$, Єю яЁртр  ўрёЄ№ (\ref{XrayRot}) -- ¤Єю
шэЄхуЁры ЇєэъЎшш $f(\vec x)$ яю яютхЁэєЄюьє ъЁєує $C'$, яюыєўхээюьє шч $C$ яЁшьхэхэшхь тЁр∙хэш  $R^{-1}_{ij}(\phi)$.
╧юф фхщёЄтшхь ¤Єюую тЁр∙хэш  ьрЄЁшЎр яы■ъъхЁют√ї ъююЁфшэрЄ ёюяЁ урхЄё  ё яюью∙№■ ьрЄЁшЎ√ тЁр∙хэш 
$R_{ij}(\phi)$. ─хщёЄтшЄхы№эю, яЁ ьюх т√ўшёыхэшх ъююЁфшэрЄ ъЁєур $C'$ фрхЄ 
(фы  ъЁрЄъюёЄш $R_{ij}(\phi)$ юсючэрўхэю чр $R$):
$$
\hat l' = x'p'^T - p'x'^T=
R^{-1} x (R^{-1} p)^T - R^{-1} p (R^{-1} x)^T =
R^{-1} (xp^T - px^T) R = R^{-1} \, \hat l \, R.
$$
╚Єръ, ь√ яюыєўшыш, ўЄю чэрўхэшх ыєўхтюую яЁхюсЁрчютрэш  $J$ юЄ яютхЁэєЄющ ЇєэъЎшш $R_{ij}(\phi) f$ эр
сюы№°юь ъЁєух ё ьрЄЁшЎхщ ъююЁфшэрЄ $\hat l$
Ёртэю чэрўхэш■ яЁхюсЁрчютрэш  юЄ шёїюфэющ ЇєэъЎшш $f$ эр ъЁєух ё ёюяЁ цхээющ
ьрЄЁшЎхщ ъююЁфшэрЄ $ R^{-1}_{ij}(\phi) \, \hat l \, R_{ij}(\phi) $:
$$
\left(\left( J \circ R_{ij}(\phi) \right) f\right) (\hat l) = (J f) (R^{-1}_{ij}(\phi) \, \hat l \, R_{ij}(\phi)).
$$
╧Ёшьхэ   яЁюшчтюфэє■ яю $\phi$ т Єюўъх $\phi=0$, яюыєўрхь:
$$
\displaystyle \left(\left(J \circ m_{ij}\right) f \right) (\hat l)
= - \left. \frac{d}{d\phi} \left(\left( J \circ R_{ij}(\phi) \right) f\right) (\hat l) \right|_{\phi = 0}
= - \left. \frac{d}{d\phi} (J f) \left(R^{-1}_{ij}(\phi) \, \hat l \, R_{ij}(\phi)\right) \right|_{\phi = 0} = \\
$$
$$
\displaystyle = - \frac12 \frac{\pd (J f)}{\pd l_{ab}}
\left. \frac{d}{d\phi} \left(R^{-1}_{ij}(\phi) \, \hat l \, R_{ij}(\phi)\right)_{ab} \right|_{\phi = 0} =
\frac12 \frac{\pd (J f)}{\pd l_{ab}}
\left( \left. \frac{d}{d\phi} R_{ij}(\phi)\right|_{\phi = 0}
\, \hat l - \hat l \,
\left. \frac{d}{d\phi} R_{ij}(\phi)\right|_{\phi = 0} \right)_{ab}  =
$$
$$
= - \frac12 \frac{\pd (J f)}{\pd l_{ab}}
\left[\hat l, \, E_{ij} \right]_{ab}  =
\frac14 \frac{\pd (J f)}{\pd l_{ab}} l_{pq}
\left[E_{pq}, \, E_{ij} \right]_{ab}  =
\frac14 \frac{\pd (J f)}{\pd l_{ab}} l_{pq} (C_{pq, ij, rs} E_{rs})_{ab} =
\frac14 \frac{\pd (J f)}{\pd l_{ab}} l_{pq} C_{pq, ij, rs} (-\delta_{ra}\delta_{sb} + \delta_{rb}\delta_{sa}) =
$$
$$
=\frac14 \frac{\pd (J f)}{\pd l_{ab}} l_{pq} (- C_{pq, ij, ab} + C_{pq, ij, ba}) =
- \frac12 \frac{\pd (J f)}{\pd l_{ab}} l_{pq} C_{pq, ij, ab} =
\frac12 \frac{\pd (J f)}{\pd l_{ab}} C_{ab, ij, pq} l_{pq} =
\frac12 \frac{\pd (J f)}{\pd l_{ab}} \left\{l_{ab}, l_{ij}\right\}  =
 \left\{J f, l_{ij}\right\}.
$$

╚Єръ, ь√ яюыєўшыш юёэютэюх ъюььєЄрЎшюээюх ёююЄэю°хэшх (\ref{comm}).
╥хяхЁ№ ь√ яЁшьхэ хь хую ъ яЁртющ ўрёЄш юёЁхфхэхээ√ї єЁртэхэшщ (\ref{momAv})
фы  ьюьхэЄр ш яюыєўрхь, ўЄю юэш шьх■Є урьшы№Єюэютє ёЄЁєъЄєЁє:
\begin{equation}
\displaystyle
\dot{l}_{ij} = \{l_{ij}, H(\hat l)\},
\label{momHamEq}
\end{equation}
уфх урьшы№Єюэшрэ хёЄ№ ыєўхтюх яЁхюсЁрчютрэшх юЄ ЇєэъЎшш фхЇюЁьрЎшш:
\begin{equation}
H(\hat l) = \frac{\varepsilon}{2\pi} J \psi.
\label{momHam1}
\end{equation}

\section{╬уЁрэшўхэшх ёшёЄхь√ эр ьэюуююсЁрчшх ├Ёрёёьрэр $G(2,n)$ ъръ эр яєрёёюэютю яюфьэюуююсЁрчшх $so(n)$}

╧ЁюёЄЁрэёЄтю $so(n)$ тёхї ъюёюёшььхЄЁшўхёъшї ьрЄЁшЎ $l_{ij}$ шьххЄ ЁрчьхЁэюёЄ№ $n(n-1)/2$.
╤ъюсър ╦ш-╧єрёёюэр (\ref{momPois}) яЁхтЁр∙рхЄ хую т яєрёёюэютю ьэюуююсЁрчшх, эр ъюЄюЁюь, тююс∙х уютюЁ ,
ш юяЁхфхыхэр ёшёЄхьр  (\ref{momHamEq}),  (\ref{momHam1}).
╬фэръю, ъръ с√ыю чрьхўхэю т√°х, ъюьяюэхэЄ√ єуыютюую ьюьхэЄр ўрёЄшЎ√ (юэш цх - яы■ъъхЁют√ ъююЁфшэрЄ√
фтєьхЁэющ яыюёъюёЄш Єхъє∙хую сюы№°юую ъЁєур) єфютыхЄтюЁ ■Є ёююЄэю°хэш ь ╧ы■ъъхЁр~(\ref{Rels}).
╧ю¤Єюьє шэЄхЁхёє■∙шх эрё ЄЁрхъЄюЁшш ёшёЄхь√ (\ref{momHamEq}) фюыцэ√ яюыэюёЄ№■ ыхцрЄ№
эр ьэюуююсЁрчшш, чрфртрхьюь ¤Єшьш ёююЄэю°хэш ьш -- ьэюуююсЁрчшш ├Ёрёёьрэр $G(2,n)$
ЁрчьхЁэюёЄш $2n-4$ (эряюьэшь, ўЄю $l_{ij}$ -- юфэюЁюфэ√х ъююЁфшэрЄ√, ш фы  юфэючэрўэюую ёююЄтхЄёЄтш 
ё яыюёъюёЄ ьш эєцэю юЄюцфхёЄтшЄ№ ьрЄЁшЎ√ тшфр $\alpha \, \hat l, \alpha \in \mathbb R$).

╧ЁшэрфыхцэюёЄ№ ЄЁрхъЄюЁшщ ьэюуююсЁрчш■ $G(2,n)$ юсхёяхўштрхЄё  ёыхфє■∙шь юс∙шь ЇръЄюь:
¤Єю ьэюуююсЁрчшх шэтрЁшрэЄэю юЄэюёшЄхы№эю ы■сющ
урьшы№Єюэютющ ёшёЄхь√ ёю ёъюсърьш (\ref{momPois}). ▌Єю ёыхфєхЄ шч Єюую, ўЄю $G(2,n)$ -- {\it яєрёёюэютю
яюфьэюуююсЁрчшх} т $so(n)$ ёю ёъюсърьш (\ref{momPois}). ╧юёыхфэхх ючэрўрхЄ, ўЄю юуЁрэшўхэшх
эр $G(2,n)$ ёъюсъш ╧єрёёюэр фтєї яЁюшчтюы№э√ї ЇєэъЎшщ эр $so(n)$ чртшёшЄ Єюы№ъю юЄ чэрўхэшщ
¤Єшї ЇєэъЎшщ эр $G(2,n)$, \cite{BMMethods}. ▌Єю, т ётю■ юўхЁхф№, ёыхфєхЄ шч ёыхфє■∙хую ётющёЄтр:
ёъюсъш ╧єрёёюэр ыхт√ї ўрёЄхщ ёююЄэю°хэшщ ╧ы■ъъхЁр $l_{j[k_1} l_{k_2 k_3]}$, чрфр■∙шї $G(2,n)$,
ёю тёхьш яхЁхьхээ√ьш $l_{ij}$ Ёртэ√ ышсю эєы■, ышсю $\pm l_{p [q_1} l_{q_2 q_3]}$
фы  эхъюЄюЁ√ї $p, q_1, q_2, q_3$.
╧юёъюы№ъє эр ьэюуююсЁрчшш ├Ёрёёьрэр $G(2,n)$ тёх ¤Єш яюышэюь√ Ёртэ√ эєы■,
Єю т ёшыє т√°хёърчрээюую юэш шьх■Є эр $G(2,n)$ эєыхт√х ёъюсъш ╧єрёёюэр ёю тёхьш яхЁхьхээ√ьш~$l_{ij}$.
╥хь ёрь√ь, юэш  ты ■Єё  шэЄхуЁрырьш фтшцхэш  фы  Єхї ЄЁрхъЄюЁшщ урьшы№Єюэют√ї ёшёЄхь, ъюЄюЁ√х
ёЄрЁЄє■Є ё $G(2,n)$, р чэрўшЄ, ¤Єш ЄЁрхъЄюЁшш юёЄр■Єё  эр $G(2,n)$ эр тёхь ётюхь яЁюЄ цхэшш.

╚Єръ, юёЁхфэхээр  ёшёЄхьр (\ref{momHamEq}) юуЁрэшўштрхЄё  эр шэтрЁшрэЄэюх яєрёёюэютю
яюфьэюуююсЁрчшх $G(2,n)$ ЁрчьхЁэюёЄш $2n-4$ т яЁюёЄЁрэёЄтх яхЁхьхээ√ї $l_{ij}, i<j$,
шьх■∙хь ЁрчьхЁэюёЄ№ $n(n-1)/2$. 

╤ Єюўъш чЁхэш  ьрЄЁшЎ√ $l_{ij}$ яЁшэрфыхцэюёЄ№ ьэюуююсЁрчш■ $G(2,n)$ ¤ътштрыхэЄэр ЁрчыюцшьюёЄш
т тшфх тэх°эхую яЁюшчтхфхэш  фтєї тхъЄюЁют, шыш, ўЄю Єю цх, Єюьє, ўЄю ьрЄЁшЎр $l_{ij}$ шьххЄ Ёрэу 2.
┬ рыухсЁх ╦ш $so(n)$ ¤Єш ьрЄЁшЎ√ юсЁрчє■Є юЁсшЄє ъюяЁшёюхфшэхээюую яЁхфёЄртыхэш  уЁєяя√ $SO(n)$,
ёь. \cite{BMMethods}.

\section{╥юяюыюуш  Ёх°хэшщ ЁхфєЎшЁютрээющ ёшёЄхь√ фы  ъюэъЁхЄэ√ї яютхЁїэюёЄхщ}

═р ¤Єрях шёёыхфютрэш  Ёрёяюыюцхэш  ухюфхчшўхёъшї эр ъюэъЁхЄэ√ї яютхЁїэюёЄ ї ъы■ўхтє■ Ёюы№ шуЁр■Є
Єюяюыюушўхёъшх ьхЄюф√. ┬ ўрёЄэюёЄш, фы  фтєьхЁэ√ї фхЇюЁьшЁютрээ√ї ёЇхЁ ё яюью∙№■ яюёЄЁюхээющ рёшьяЄюЄшўхёъющ ЁхфєъЎшш
єфрхЄё  юёє∙хёЄтшЄ№ яюыэ√щ Єюяюыюушўхёъшщ рэрышч ьэюцхёЄтр ухюфхчшўхёъшї.
╩Ёюьх Єюую, фы  ЄЁхїьхЁэ√ї фхЇюЁьшЁютрээ√ї ёЇхЁ тЁр∙хэш  ЁхфєЎшЁютрээр  ёшёЄхьр юърч√трхЄё  шэЄхуЁшЁєхьющ
ёшёЄхьющ ё фтєь  ёЄхяхэ ьш ётюсюф√ ш фюяєёърхЄ шёёыхфютрэшх ьхЄюфрьш Єюяюыюушўхёъющ ъырёёшЇшърЎшш шэЄхуЁшЁєхь√ї ёшёЄхь.
╨рёёьюЄЁшь яюфЁюсэхх ёююЄтхЄёЄтє■∙шх яЁшьхЁ√ яютхЁїэюёЄхщ т ърцфюь шч ¤Єшї ъырёёют.

\subsection{─тєьхЁэ√х фхЇюЁьшЁютрээ√х ёЇхЁ√}

╨рёёьюЄЁшь ёыєўрщ $n=3$, Є.х. ёыєўрщ фтєьхЁэющ фхЇюЁьшЁютрээющ ёЇхЁ√ т ЄЁхїьхЁэюь хтъышфютюь яЁюёЄЁрэёЄтх,
ш тюч№ьхь фхЇюЁьрЎш■ т тшфх ёєьь√ ўхЄтхЁЄ√ї ёЄхяхэхщ ъююЁфшэрЄ ё Ёрчышўэ√ьш ъю¤ЇЇшЎшхэЄрьш:
\begin{equation}
  \varphi(\vec x) \, \equiv \, x_1^2 + x_2^2 + x_3^2 - 1 + \varepsilon \,\psi(\vec x) = 0, \quad \varepsilon \ll 1, \quad
  \psi(\vec x) = \varepsilon_1 x_1^4 + \varepsilon_2 x_2^4 + \varepsilon_3 x_3^4.
\label{Deg4}
\end{equation}
╙уыютющ ьюьхэЄ т ЄЁхїьхЁэюь яЁюёЄЁрэёЄтх шьххЄ ЄЁш ёє∙хёЄтхээ√ї ъюьяюэхэЄ√: $l_{12}, l_{13}, l_{23}$.
┬ ¤Єюь ёыєўрх єфюсэю ттхёЄш, ъръ ¤Єю юс√ўэю ш фхыр■Є, тьхёЄю ъюёюёшььхЄЁшўхёъющ ьрЄЁшЎ√ $l_{ij}$
ЄЁхїьхЁэ√щ тхъЄюЁ єуыютюую ьюьхэЄр $\vec L$ яю ЇюЁьєых: $L_i = \varepsilon_{ijk} l_{jk}$, ўЄю т  тэюь тшфх т√уы фшЄ ъръ:
$$
\vec L = (L_1, L_2, L_3) = (l_{23}, -l_{13}, l_{12}).
$$
╤ъюсъш ╧єрёёюэр фы  ъюьяюэхэЄ тхъЄюЁр ьюьхэЄр шьх■Є тшф:
\begin{equation}
\{L_i, L_j\} = \sum_{k=1}^3 \varepsilon_{ijk} L_k.
\label{Poiss3D}
\end{equation}
╤ъюсъш (\ref{Poiss3D}) шьх■Є юфэє ЇєэъЎш■ ╩рчшьшЁр -- ътрфЁрЄ ьюьхэЄр $\vec L^2 = L_1^2 + L_2^2 + L_3^2$.
╘шъёшЁє  хх чэрўхэшх, яюыєўрхь, ўЄю ЁхфєЎшЁютрээє■ ёшёЄхьє ьюцэю юуЁрэшўшЄ№ эр ёЇхЁє $\vec L^2 = const$.
╩Ёюьх Єюую, яЁюяюЁЎшюэры№э√х тхъЄюЁ√ ьюьхэЄр ёююЄтхЄёЄтє■Є юфэюьє ш Єюьє цх сюы№°юьє ъЁєує,
яю¤Єюьє ьэюцхёЄтю сюы№°шї ъЁєуют (шыш ёюфхЁцр∙шї шї фтєьхЁэ√ї яыюёъюёЄхщ, яЁюїюф ∙шї ўхЁхч ЎхэЄЁ)
ёююЄтхЄёЄтєхЄ ьэюцхёЄтє яЁ ь√ї т яЁюёЄЁрэёЄтх єуыютюую ьюьхэЄр, яЁюїюф ∙шї ЎхЁхч эрўрыю ъююЁфшэрЄ,
Є.х. яЁюхъЄштэющ яыюёъюёЄш $\mathbb {R P}^2$. ╚Єръ, ЁхфєЎшЁютрээр  ёшёЄхьр юяЁхфхыхэр эр уЁрёёьрэшрэх $G(2,3)$,
уюьхюьюЁЇэюь яЁюхъЄштэющ яыюёъюёЄш.

┬ ёююЄтхЄёЄтшш ё юс∙хщ ёїхьющ (\ref{momHamEq}), (\ref{momHam1}) урьшы№Єюэшрэ ЁхфєЎшЁютрээющ ёшёЄхь√ яюыєўрхЄё 
 шч ЇєэъЎшш $\psi(\vec x)$ ё яюью∙№■ ыєўхтюую яЁхюсЁрчютрэш , ъюЄюЁюх т ёыєўрх фтєьхЁэющ ёЇхЁ√
 эрч√тр■Є яЁхюсЁрчютрэшхь ╘єэър-╠шэъютёъюую.
─ы  ЁрёёьрЄЁштрхьюую тючьє∙хэш  ўхЄтхЁЄ√ьш ёЄхяхэ ьш (\ref{Deg4}) урьшы№Єюэшрэ шьххЄ тшф:
\begin{equation}
H = \frac38 \, \varepsilon \left[ \varepsilon_1 \,(L_2^2 + L_3^2)^2 + \varepsilon_2 \,(L_1^2 + L_3^2)^2 + \varepsilon_3 \,(L_1^2 + L_2^2)^2 \right].
\label{HamDeg4}
\end{equation}
╧юёъюы№ъє Їрчютюх яЁюёЄЁрэёЄтю ЁхфєЎшЁютрээющ ёшёЄхь√ фтєьхЁэю, р ёшёЄхьр урьшы№Єюэютр, Єю юэр тяюыэх шэЄхуЁшЁєхьр.
┼х ЄЁрхъЄюЁш ьш  ты ■Єё  ышэшш єЁютэ  урьшы№Єюэшрэр $H = const$ эр $\mathbb {R P}^2$.
╥юяюыюуш■ шї Ёрёяюыюцхэш  ьюцэю шчєўшЄ№ яЁш яюью∙ш Їрчютюую яюЁЄЁхЄр.
─ы  ¤Єюую юяЁхфхы ■Єё  ёЄрЎшюэрЁэ√х Єюўъш ёшёЄхь√, т ъюЄюЁ√ї $\dot{L}_i = 0$.
╬эш ёююЄтхЄёЄтє■Є ухюфхчшўхёъшь, ыхцр∙шь тсышчш юфэюую сюы№°юую ъЁєур.
─рыхх, шёёыхфєхЄё  єёЄющўштюёЄ№ ¤Єшї ёЄрЎшюэрЁэ√ї Єюўхъ, т Ёхчєы№ЄрЄх ўхую юэш яюфЁрчфхы ■Єё  эр ЎхэЄЁ√
ш ёхфыр. ╤хфыют√х Єюўъш ёюхфшэ ■Єё  ёхярЁрЄЁшёрьш.

┬ чртшёшьюёЄш юЄ ярЁрьхЄЁют $\varepsilon_1, \varepsilon_2, \varepsilon_3$ урьшы№Єюэшрэ (\ref{HamDeg4})
ьюцхЄ шьхЄ№ ёЄрЎшюэрЁэ√х Єюўъш ёыхфє■∙шї ЄЁхї тшфют:

\begin{itemize}
  \item[\sf S1] \label{S1}
    ╤є∙хёЄтє■Є яЁш ы■с√ї чэрўхэш ї ярЁрьхЄЁют $\varepsilon_1, \varepsilon_2, \varepsilon_3$.
    ╟эрўхэш  ьюьхэЄр т ёЄрЎшюэрЁэ√ї Єюўърї єфютыхЄтюЁ ■Є єёыютш ь:
    \begin{itemize}
    \item[\sf a.] \label{T1a}
    $ L_{10}  =  0, \quad L_{20}  =  0, \quad L_{30} \ne 0; $
    \item[\sf b.] \label{T1b}
        $ L_{10}  =  0, \quad L_{20} \ne 0, \quad L_{30}  =  0; $
        \item[\sf c.] \label{T1c}
     $ L_{10} \ne 0, \quad L_{20}  =  0, \quad L_{30}  =  0; $
    \end{itemize}
  \item[\sf S2] \label{S2}
    ╤є∙хёЄтє■Є яЁш єёыютш ї $\varepsilon_2 \varepsilon_3 > 0$, $\varepsilon_3 \varepsilon_1 > 0$, ш $\varepsilon_1
    \varepsilon_2 > 0$ ёююЄтхЄёЄтхээю.
    ╟эрўхэш  ьюьхэЄр т ёЄрЎшюэрЁэ√ї Єюўърї єфютыхЄтюЁ ■Є єёыютш ь:
    \begin{itemize}
    \item[\sf a.] \label{T2a}
    $ L_{10} = 0, \quad L_{20} \ne 0, \quad L_{30} \ne 0, \quad
      \varepsilon_3 L_{20}^2 - \varepsilon_2 L_{30}^2 = 0; $
    \item[\sf b.] \label{T2b}
      $ L_{20} = 0, \quad L_{30} \ne 0, \quad L_{10} \ne 0, \quad
       \varepsilon_1 L_{30}^2 - \varepsilon_3 L_{10}^2 = 0; $
    \item[\sf c.] \label{T2c}
     $L_{30} = 0, \quad L_{10} \ne 0, \quad L_{20} \ne 0, \quad
     \varepsilon_2 L_{10}^2 - \varepsilon_1 L_{20}^2 = 0; $
    \end{itemize}
  \item[\sf S3] \label{S3}
    ╤є∙хёЄтє■Є яЁш єёыютш ї:
        \begin{eqnarray}
            \varepsilon_1 \,\varepsilon_2 - \varepsilon_2 \,\varepsilon_3 +
            \varepsilon_3
            \,\varepsilon_1  \, > 0  \nonumber  \\
            \varepsilon_1 \,\varepsilon_2 + \varepsilon_2 \,\varepsilon_3 -
            \varepsilon_3
            \,\varepsilon_1  \, > 0  \nonumber  \\
            \ - \varepsilon_1 \,\varepsilon_2 + \varepsilon_2 \,\varepsilon_3
            + \varepsilon_3
            \,\varepsilon_1  \, > 0  \nonumber
        \end{eqnarray}
    ╟эрўхэш  ьюьхэЄр т ёЄрЎшюэрЁэ√ї Єюўърї єфютыхЄтюЁ ■Є єёыютш ь:
    $L_{10} \ne 0, L_{20} \ne 0,  L_{30} \ne 0$ ш
    $$
        \frac{L_{10}^2}
        {\varepsilon_1 \,\varepsilon_2 - \varepsilon_2 \,
        \varepsilon_3 + \varepsilon_3
        \,\varepsilon_1} =
        \frac{L_{20}^2}
        {\varepsilon_1 \,\varepsilon_2 + \varepsilon_2
        \,\varepsilon_3 - \varepsilon_3
        \,\varepsilon_1} =
        \frac{L_{30}^2}
        { - \varepsilon_1 \,\varepsilon_2 + \varepsilon_2
        \,\varepsilon_3 + \varepsilon_3
        \,\varepsilon_1}
    $$
 \end{itemize}
╤ яюью∙№■ ышэхрЁшчрЎшш єЁртэхэшщ фтшцхэш  т юъЁхёЄэюёЄш ¤Єшї Єюўхъ ш яюфёўхЄр ёюсёЄтхээ√ї чэрўхэшщ
эрїюфшь, ўЄю ¤Єш ёЄрЎшюэрЁэ√х Єюўъш  ты ■Єё  ЎхэЄЁрьш яЁш ёыхфє■∙шї єёыютш ї (ёююЄтхЄёЄтхээю):
\begin{itemize}
    \item[\sf S1]
    \begin{itemize}
        \item[\sf a.] $\varepsilon_1 \varepsilon_2 > 0$;
        \item[\sf b.] $\varepsilon_2 \varepsilon_3 > 0$;
        \item[\sf c.] $\varepsilon_3 \varepsilon_1 > 0$.
    \end{itemize}
    \item[\sf S2]
    \begin{itemize}
        \item[\sf a.] $ \varepsilon_1 \varepsilon_2
            -\varepsilon_2 \varepsilon_3
            +\varepsilon_3 \varepsilon_1 < 0 $;
        \item[\sf b.] $ \varepsilon_1 \varepsilon_2
            +\varepsilon_2  \varepsilon_3
            -\varepsilon_3 \varepsilon_1 < 0 $;
        \item[\sf c.] $-\varepsilon_1 \varepsilon_2
            +\varepsilon_2 \varepsilon_3
            +\varepsilon_3 \varepsilon_1 < 0 $.
    \end{itemize}
    \item[\sf S3] \quad ╧Ёш ы■с√ї $ \varepsilon_i $.
\end{itemize}
╥хь ёрь√ь фы  ы■сюую эрсюЁр ярЁрьхЄЁют яютхЁїэюёЄш $ \varepsilon_i $ т√ўшёы ■Єё  ёЄрЎшюэрЁэ√х Єюўъш
ш т√ ёэ хЄё  шї Ёрчфхыхэшх эр ЎхэЄЁ√ ш ёхфыр.
╚ч рэрышчр т√°хяЁштхфхээ√ї єёыютшщ эр $ \varepsilon_i $ ёыхфєхЄ, ўЄю ёє∙хёЄтє■Є ёыхфє■∙шх 4 Єюяюыюушўхёъш Ёрчышўэ√ї
Єшяр Їрчют√ї яюЁЄЁхЄют ЁхфєЎшЁютрээющ ёшёЄхь√:
\begin{itemize}
  \item[\sf Type I] \label{T1}
        \quad  7 ЎхэЄЁют ш 6 ёхфхы;
        $ \varepsilon_i $ яюфўшэхэ√ єёыютш ь:
      \begin{equation}
      \begin{array}{rcl}
     \varepsilon_1 \varepsilon_2
    -\varepsilon_2 \varepsilon_3
    +\varepsilon_3 \varepsilon_1 > 0 \\

     \varepsilon_1 \varepsilon_2
    +\varepsilon_2 \varepsilon_3
    -\varepsilon_3 \varepsilon_1 > 0 \\

    -\varepsilon_1 \varepsilon_2
    +\varepsilon_2 \varepsilon_3
    +\varepsilon_3 \varepsilon_1 > 0
      \end{array}
      \mbox{.}
     \label{t1constraints}
      \end{equation}

 \item[\sf Type II] \label{T2}
       \quad 5 ЎхэЄЁют ш 4 ёхфыр;
       $ \varepsilon_i $ эх Ёртэ√ эєы■, шьх■Є юфшэ чэръ ш їюЄ  с√ юфэю шч єёыютшщ (\ref{t1constraints}) эх т√яюыэхэю.

 \item[\sf Type III] \label{T3}
       \quad 3 ЎхэЄЁр ш 2 ёхфыр;
       $ \varepsilon_i $ єфютыхЄтюЁ ■Є юфэюьє шч ёыхфє■∙шї єёыютшщ:
       $ \varepsilon_2 \varepsilon_3 > 0 $ ш
       $ \varepsilon_1 \varepsilon_2 \leq 0 $; \quad
       $ \varepsilon_3 \varepsilon_1 > 0 $ ш
       $ \varepsilon_2 \varepsilon_3 \leq 0 $; \quad
       $ \varepsilon_1 \varepsilon_2 > 0  $ ш
       $ \varepsilon_3 \varepsilon_1 \leq 0 $.

 \item[\sf Type IV] \label{T4}
      \quad 2 ЎхэЄЁр ш 1 ёхфыю;
       $ \varepsilon_i $ єфютыхЄтюЁ ■Є юфэюьє шч ёыхфє■∙шї єёыютшщ:
       $ \varepsilon_1 = 0 $ ш
       $ \varepsilon_2 \varepsilon_3 \leq 0 $; \quad
       $ \varepsilon_2 = 0 $ ш
       $ \varepsilon_3 \varepsilon_1 \leq 0 $; \quad
       $ \varepsilon_3 = 0 $ ш
       $ \varepsilon_1 \varepsilon_2 \leq 0 $.

  \end{itemize}

\begin{figure}
  \begin{center}
    \includegraphics[width = 500bp]{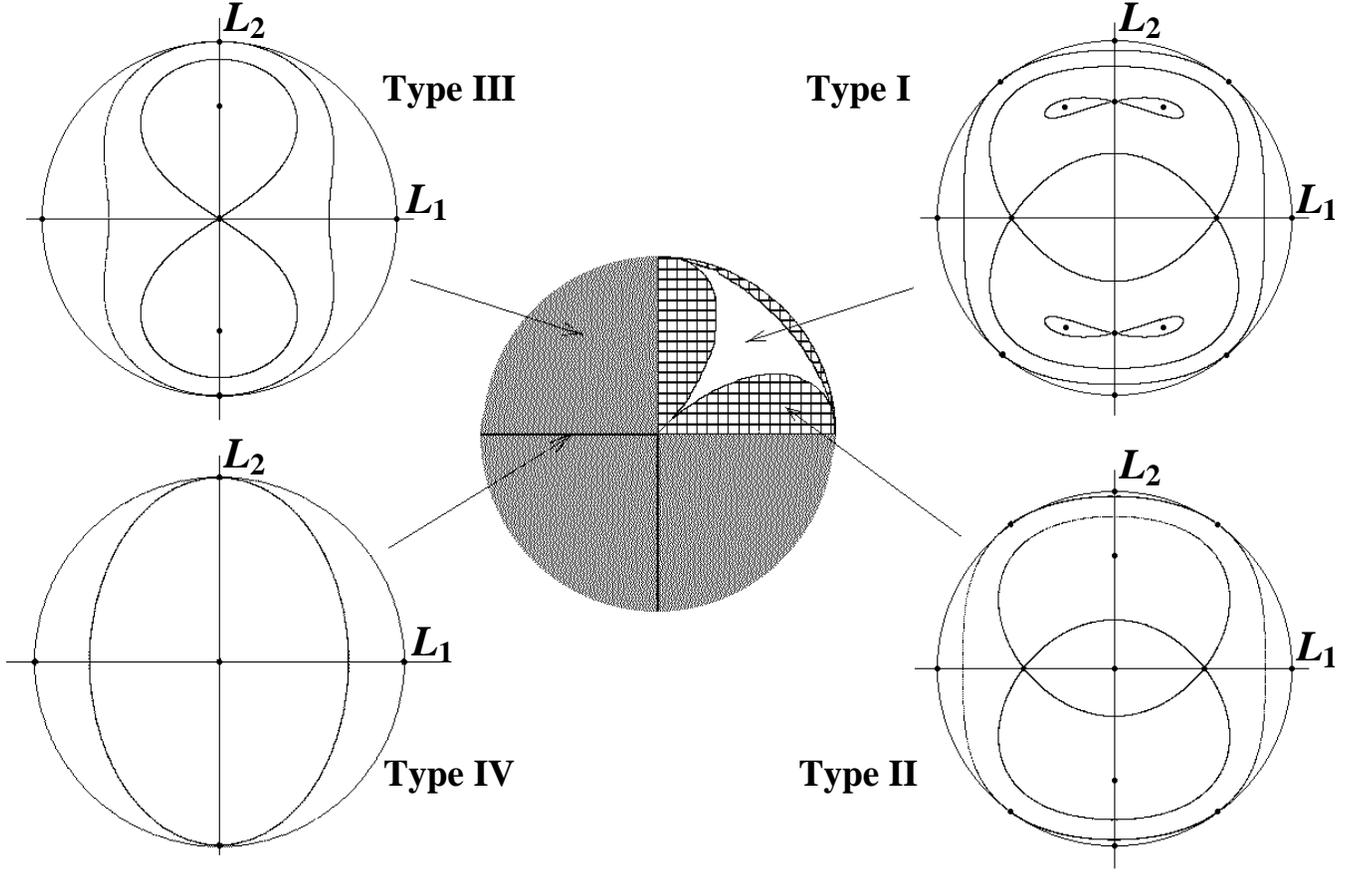}
    \caption{
     ╬сырёЄш т яЁюёЄЁрэёЄтх ярЁрьхЄЁют $\varepsilon_i$, ёююЄтхЄёЄтє■∙шх Єшярь I - IV
     Їрчют√ї яюЁЄЁхЄют ЁхфєЎшЁютрээющ ёшёЄхь√.
     ╦шэшш, Ёрчфхы ■∙шх юсырёЄш Єшяют I ш II чрфр■Єё  єЁртэхэш ьш (\ref{boundary}).
    }
    \label{fig2}
  \end{center}
\end{figure}

┬тшфє юфэюЁюфэюёЄш яЁрт√ї ўрёЄхщ ЁхфєЎшЁютрээющ ёшёЄхь√ (\ref{momHamEq}) яю $ \varepsilon_i $
єьэюцхэшх тёхї $ \varepsilon_i $ эр юфэю ўшёыю эх ьхэ хЄ Їрчют√щ яюЁЄЁхЄ.
╧ю¤Єюьє ьюцэю ЁрёёьрЄЁштрЄ№ ъюьсшэрЎшш ¤Єшї ярЁрьхЄЁют ё ЇшъёшЁютрээющ ёєььющ ътрфЁрЄют, Є.х. ыхцр∙шх эр ёЇхЁх,
ш яЁш ¤Єюь фюёЄрЄюўэю ЁрёёьюЄЁхЄ№ юфэє яюыєёЇхЁє.
═р Ёшё. \ref{fig2} шчюсЁрцхэ√ юсырёЄш эр яюыєёЇхЁх т яЁюёЄЁрэёЄтх ярЁрьхЄЁют $ \varepsilon_i $, ёююЄтхЄёЄтє■∙шх
фхЇюЁьшЁютрээ√ь ёЇхЁрь, фы  ъюЄюЁ√ї ЁхфєЎшЁютрээр  ёшёЄхьр шьххЄ ърцф√щ шч яЁштхфхээ√ї ўхЄ√Ёхї Єшяют Їрчют√ї яюЁЄЁхЄют.
╬сырёЄш, ёююЄтхЄёЄтє■∙шх Єшярь I ш II, Ёрчфхы ■Єё  ышэш ьш, чрфрээ√ьш юфэюЁюфэ√ьш єЁртэхэш ьш тЄюЁюую яюЁ фър:
\begin{eqnarray}
   &1.&\quad \varepsilon_1 \varepsilon_2
  -\varepsilon_2 \varepsilon_3
  +\varepsilon_3 \varepsilon_1 = 0 \label{boundary} \\
   &2.&\quad \varepsilon_1 \varepsilon_2
  +\varepsilon_2 \varepsilon_3
  -\varepsilon_3 \varepsilon_1 = 0 \nonumber \\
   &3.&\quad -\varepsilon_1 \varepsilon_2
  +\varepsilon_2 \varepsilon_3
  +\varepsilon_3 \varepsilon_1 = 0 \nonumber
\end{eqnarray}

╚Єръ, т ёыєўрх фтєьхЁэ√ї фхЇюЁьшЁютрээ√ї ёЇхЁ яюёЄЁюхээр  рёшьяЄюЄшўхёър  урьшы№Єюэютр ЁхфєъЎш 
яючтюы хЄ яЁютхёЄш яюыэ√щ Єюяюыюушўхёъшщ рэрышч ёютюъєяэюёЄш ухюфхчшўхёъшї эр яютхЁїэюёЄш
яєЄхь яюёЄЁюхэш  Їрчютюую яюЁЄЁхЄр ЁхфєЎшЁютрээющ ёшёЄхь√.
┬ чртшёшьюёЄш юЄ ярЁрьхЄЁют фхЇюЁьрЎшш яЁюшёїюф Є Єюяюыюушўхёъшх яхЁхёЄЁющъш Їрчют√ї яюЁЄЁхЄют.

\subsection{╥ЁхїьхЁэ√х фхЇюЁьшЁютрээ√х ёЇхЁ√}

╨рёёьюЄЁшь ёыєўрщ $n=4$, Є.х. ухюфхчшўхёъшх эр ЄЁхїьхЁэ√ї фхЇюЁьшЁютрээ√ї ёЇхЁрї т ўхЄ√ЁхїьхЁэюь хтъышфютюь яЁюёЄЁрэёЄтх.
─хЇюЁьрЎш■, рэрыюушўэю фтєьхЁэюьє ёыєўр■, т√схЁхь т тшфх ёєьь√ ўхЄтхЁЄ√ї ёЄхяхэхщ ъююЁфшэрЄ:
\begin{equation}
  \varphi(\vec x) \, \equiv \, x_1^2 + x_2^2 + x_3^2 + x_4^2 - 1 + \varepsilon \,\psi(\vec x) = 0, \quad \varepsilon \ll 1, \quad
  \psi(\vec x) = \varepsilon_1 x_1^4 + \varepsilon_2 x_2^4 + \varepsilon_3 x_3^4 + \varepsilon_4 x_4^4.
\label{Deg4S3}
\end{equation}
╙уыютющ ьюьхэЄ т ўхЄ√ЁхїьхЁэюь яЁюёЄЁрэёЄтх шьххЄ °хёЄ№ ёє∙хёЄтхээ√ї ъюьяюэхэЄ: $l_{12}, l_{13}, l_{14}, l_{23}, l_{24}, l_{34}$.
╬эш ёт чрэ√ юфэшь ёююЄэю°хэшхь ╧ы■ъъхЁр:
\begin{equation}
C \, \equiv \, l_{12} l_{34} - l_{13} l_{24} + l_{14} l_{23} = 0.
\label{plS3}
\end{equation}
└ыухсЁр ╧єрёёюэр єуыютюую ьюьхэЄр шьххЄ фтх ЇєэъЎшш ╩рчшьшЁр: яЁштхфхээє■ т√°х ыхтє■ ўрёЄ№ $C$ ёююЄэю°хэш  ╧ы■ъъхЁр,
р Єръцх ёєььє ътрфЁрЄют ъюьяюэхэЄ ьюьхэЄр: $l^2 = \sum_{ij} l_{ij}^2$. ╟рЇшъёшЁютрт, $C=0$ ш $l^2=1$,
р Єръцх юЄюцфхёЄты   Єюўъш, юЄышўр■∙шхё  чрьхэющ чэрър є тёхї $ l_{ij}$,
яюыєўрхь ўхЄ√ЁхїьхЁэюх ьэюуююсЁрчшх ├Ёрёёьрэр $G(2,4)$ т ърўхёЄтх Їрчютюую яЁюёЄЁрэёЄтр (ёыхфютрЄхы№эю, шьххЄё  фтх ёЄхяхэш ётюсюф√).

├рьшы№Єюэшрэ  ЁхфєЎшЁютрээющ ёшёЄхь√, ЁрёёўшЄрээ√щ яю юс∙хщ ёїхьх (\ref{momHamEq}), (\ref{momHam1}), 
фы  ЁрёёьрЄЁштрхьюую тючьє∙хэш  ўхЄтхЁЄ√ьш ёЄхяхэ ьш (\ref{Deg4S3}) шьххЄ тшф:
\begin{equation}
H = \frac38 \, \varepsilon \left[ \varepsilon_1 \,(l_{12}^2 + l_{13}^2 + l_{14}^2)^2 + \varepsilon_2 \,(l_{12}^2 + l_{23}^2 + l_{24}^2)^2
                      + \varepsilon_3 \,(l_{13}^2 + l_{23}^2 + l_{34}^2)^2 + \varepsilon_4 \,(l_{14}^2 + l_{24}^2 + l_{34}^2)^2\right].
\label{HamDeg4S3}
\end{equation}
╧юёъюы№ъє ЁхфєЎшЁютрээр  ёшёЄхьр шьххЄ фтх ёЄхяхэш ётюсюф√, Єю
фы  хх шэЄхуЁшЁєхьюёЄш ЄЁхсєхЄё  юфшэ фюяюыэшЄхы№э√щ яхЁт√щ шэЄхуЁры, яюьшью урьшы№Єюэшрэр.
╨рёёьюЄЁшь ёыєўрщ юёхтющ ёшььхЄЁшш, ъюуфр $\varepsilon_3 = \varepsilon_4 = 0$ ш
ЇєэъЎш  фхЇюЁьрЎшш шьххЄ тшф: $ \psi(\vec x) = \varepsilon_1 x_1^4 + \varepsilon_2 x_2^4 $.
╤ююЄтхЄёЄтє■∙р  фхЇюЁьшЁютрээр  ёЇхЁр  ты хЄё  яютхЁїэюёЄ№■ тЁр∙хэш : яютюЁюЄ√ т яыюёъюёЄш $x_3 x_4$ яхЁхтюф Є хх т ёхс .
╚ч ¤Єющ ёшььхЄЁшш ёыхфєхЄ Єюўэюх ёюїЁрэхэшх ъюьяюэхэЄ√ єуыютюую ьюьхэЄр $l_{34}$ яЁш фтшцхэшш ўрёЄшЎ√ яю ы■сющ ухюфхчшўхёъющ.
▌Єю цх ётющёЄтю яхЁхэюёшЄё  ш эр ЁхфєЎшЁютрээє■ ёшёЄхьє. ─хщёЄтшЄхы№эю, хх урьшы№Єюэшрэ:
$$
H = \frac38 \left[ \varepsilon_1 \,(l_{12}^2 + l_{13}^2 + l_{14}^2)^2 + \varepsilon_2 \,(l_{12}^2 + l_{23}^2 + l_{24}^2)^2 \right]
$$
шьххЄ эєыхтє■ ёъюсъє ╧єрёёюэр ё тхышўшэющ $l_{34}$, ъюЄюЁр  Єхь ёрь√ь яЁхфёЄрты хЄ ёюсющ эхфюёЄр■∙шщ яхЁт√щ шэЄхуЁры.
╧ю ЄхюЁхьх ╦шєтшыы  яюыєўрхь, ўЄю фшэрьшър ЁхфєЎшЁютрээющ ёшёЄхь√ ёюёЄюшЄ т єёыютэю-яхЁшюфшўхёъюь фтшцхэшш яю фтєьхЁэ√ь ЄюЁрь,
 ты ■∙шьё  ёютьхёЄэ√ьш яютхЁїэюёЄ ьш єЁютэ  шэЄхуЁрыют фтшцхэш , Є.х. чрфртрхь√ь єЁртэхэш ьш: $C=0, \ l^2=1, \ H = c_1, \ l_{34} = c_2$
фы  яЁюшчтюы№э√ї $c_1, c_2$.

\begin{figure}
  \begin{center}
    \includegraphics[width = 400bp]{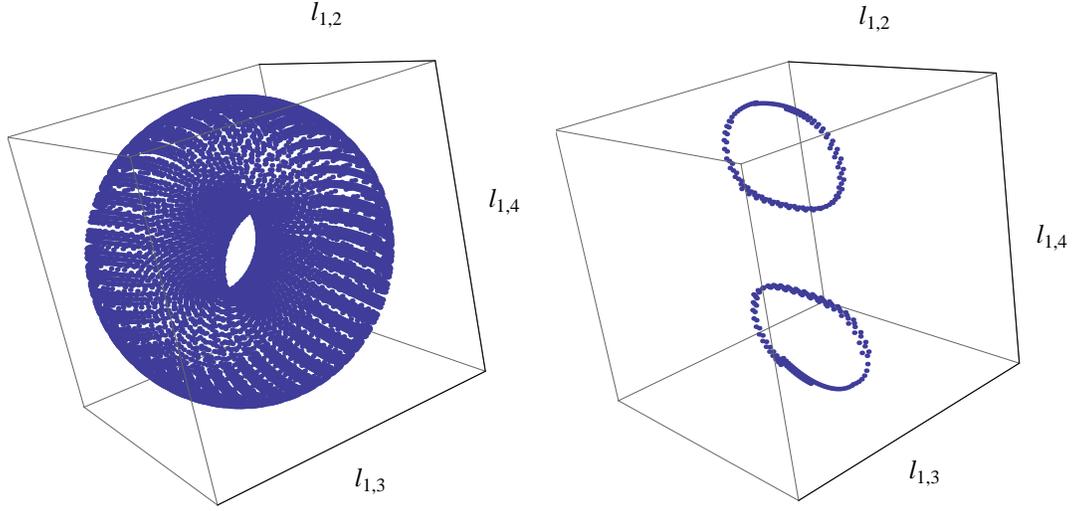}
    \caption{
       ╥ЁрхъЄюЁш  ЁхфєЎшЁютрээющ ёшёЄхь√ фы  ЄЁхїьхЁэющ ёЇхЁ√ ё фхЇюЁьрЎшхщ тшфр:
       $ \psi(\vec x) = \varepsilon_1 x_1^4 + \varepsilon_2 x_2^4 $
       (ёыхтр) ш ёхўхэшх ¤Єющ ЄЁрхъЄюЁшш яыюёъюёЄ№■ (ёяЁртр). ╧ютхЁїэюёЄ№ шэтрЁшрэЄэр юЄэюёшЄхы№эю
       тЁр∙хэшщ т яыюёъюёЄш $x_3 x_4$, ёыхфютрЄхы№эю, шьххЄё  фюяюыэшЄхы№э√щ шэЄхуЁры $l_{34}$ ш ЁхфєЎшЁютрээр  ёшёЄхьр шэЄхуЁшЁєхьр. 
       ┬ ёююЄтхЄёЄтшш ё ЄхюЁхьющ ╦шєтшыы , ЄЁрхъЄюЁш  чрьхЄрхЄ фтєьхЁэ√щ ЄюЁ.
    }
    \label{fig3}
  \end{center}
\end{figure}

═р Ёшё. \ref{fig3} шчюсЁрцхэр юфэр шч ЄЁрхъЄюЁшщ ЁхфєЎшЁютрээющ ёшёЄхь√ т ЁрёёьрЄЁштрхьюь ёыєўрх юёхтющ ёшььхЄЁшш.
┬шфэю, ўЄю юэр чрьхЄрхЄ фтєьхЁэ√щ ЄюЁ.
╚чєўхэшх Ёрёяюыюцхэш  ЄюЁют ╦шєтшыы  т Їрчютюь яЁюёЄЁрэёЄтх ьхЄюфрьш Єюяюыюушўхёъющ ъырёёшЇшърЎшш
шэЄхуЁшЁєхь√ї ёшёЄхь сєфхЄ яЁхфьхЄюь фры№эхщ°хую шёёыхфютрэш .

\begin{figure}
  \begin{center}
    \includegraphics[width = 400bp]{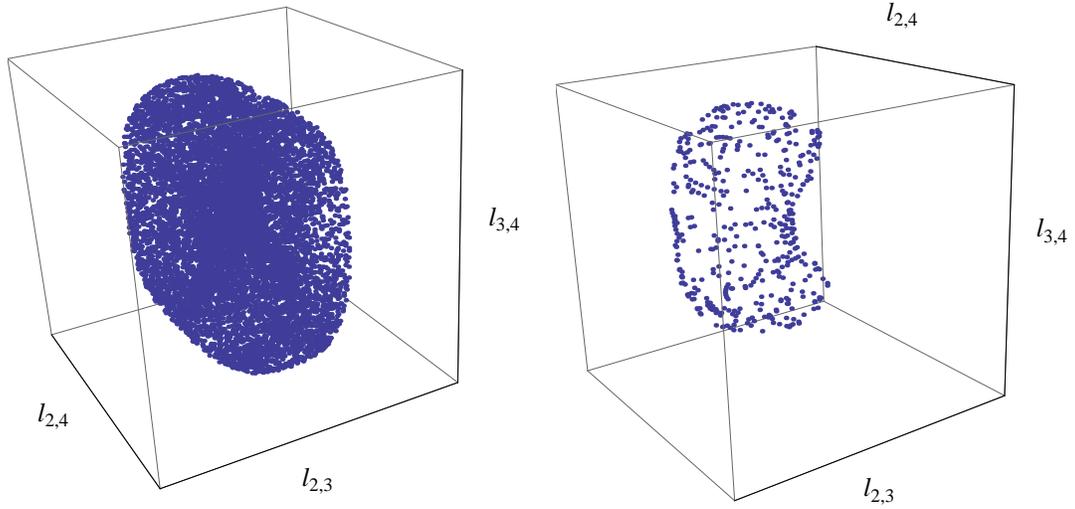}
    \caption{
       ╥ЁрхъЄюЁш  ЁхфєЎшЁютрээющ ёшёЄхь√ фы  ЄЁхїьхЁэющ ёЇхЁ√ ё фхЇюЁьрЎшхщ тшфр: 
       $ \psi(\vec x) = \varepsilon_1 x_1^4 + \varepsilon_2 x_2^4  + \varepsilon_3 x_3^4 + \varepsilon_4 x_4^4$
       (ёыхтр) ш ёхўхэшх ¤Єющ ЄЁрхъЄюЁшш яыюёъюёЄ№■ (ёяЁртр). ╥ЁрхъЄюЁш  чрьхЄрхЄ юсырёЄ№ т ЄЁхїьхЁэюь шчю¤эхЁухЄшўхёъюь
       ьэюуююсЁрчшш, ш фшэрьшър, яю-тшфьюьє, эхЁхуєы Ёэр.
    }
    \label{fig4}
  \end{center}
\end{figure}

┬ юс∙хь ёыєўрх фхЇюЁьрЎшш ўхЄтхЁЄ√ьш ёЄхяхэ ьш (\ref{Deg4S3}), ъюуфр тёх ъю¤ЇЇшЎшхэЄ√ $ \varepsilon_i $
юЄышўэ√ юЄ эєы , ьюфхышЁютрэшх ЁхфєЎшЁютрээющ ёшёЄхь√ яюърч√трхЄ, ўЄю шьх■Єё  єёыютш , ъюуфр фшэрьшър,
яю-тшфшьюьє, шьххЄ эхЁхуєы Ёэ√щ їрЁръЄхЁ,
ш ЄЁрхъЄюЁш  чрьхЄрхЄ юсырёЄ№ т ЄЁхїьхЁэюь ьэюуююсЁрчшш, чрфртрхьюь яюёЄю эёЄтюь ¤эхЁушш, Ёшё. \ref{fig4}.

\subsection{╠эюуюьхЁэ√х ¤ыышяёюшф√, сышчъшх ъ ёЇхЁх, ш ёыєўрщ ╪юЄЄъш-╠рэръютр т єЁртэхэш ї ▌щыхЁр эр рыухсЁх ╦ш $so(n)$.}

╨рёёьюЄЁшь ёыєўрщ $(n-1)$-ьхЁэюую ¤ыышяёюшфр, сышчъюую ъ ёЇхЁх, ўЄю ёююЄтхЄёЄтєхЄ ътрфЁрЄшўэющ ЇєэъЎшш фхЇюЁьрЎшш $\psi(\vec x)$:
$$
    \varphi \; \equiv \; x_1^2 \; + \; x_2^2 \; + \; \ldots \; + \; x_n^2 \; -1 \;
 + \; \varepsilon \; (\alpha_1 x_1^2 \; + \; \alpha_2 x_2^2 \; + \; \ldots \; + \; \alpha_n x_n^2) = 0.
$$
├рьшы№Єюэшрэ  ЁхфєЎшЁютрээющ ёшёЄхь√ яю ЇюЁьєых (\ref{momHam1}) юърч√трхЄё  Ёртхэ:
\begin{equation}
H = \frac12 \, \varepsilon \sum_{i<j} (\alpha_i + \alpha_j) \, l_{ij}^2.
\label{HamEll}
\end{equation}
═хЄЁєфэю тшфхЄ№, ўЄю ¤Єю ўрёЄэ√щ ёыєўрщ ьэюуюьхЁэюую шэЄхуЁшЁєхьюую ёыєўр  ╪юЄЄъш-╠рэръютр
т єЁртэхэш ї ▌щыхЁр эр рыухсЁх ╦ш $so(n)$ \cite{Schottky}, \cite{Manakov}:
\begin{equation}
H = \sum_{i<j} \frac{a_i - a_j}{b_i - b_j} \, l_{ij}^2
\label{HamEll}
\end{equation}
яЁш $a_i = \alpha_i^2, \, b_i = 2 \alpha_i$.

╥хь ёрь√ь єёЄрэртыштрхЄё  шчюьюЁЇшчь ёшёЄхь√ яхЁтюую яЁшсышцхэш  фы  чрфрўш ю ухюфхчшўхёъшї эр $(n-1)$-ьхЁэюь ¤ыышяёюшфх,
сышчъюь ъ ёЇхЁх, ё ёшёЄхьющ ╪юЄЄъш-╠рэръютр эр рыухсЁх $so(n)$.

╟фхё№ яЁхфёЄрты хЄ шэЄхЁхё ёыхфє■∙хх чрьхўрэшх. ┬√°х єяюьшэрыё  Ёхчєы№ЄрЄ ┬.┬. ╩ючыютр, └.┬. ┴юЁшёютр, ╚.╤. ╠рьрхтр, └.╠. ╧хЁхыюьютр
ю ¤ътштрыхэЄэюёЄш чрфрўш ю ухюфхчшўхёъшї эр $(n-1)$-ьхЁэюь ¤ыышяёюшфх ёыєўр■ ╩ыхс°р-╧хЁхыюьютр фы  єЁртэхэшщ ╩шЁїуюЇр
эр рыухсЁх ╦ш $e(n)$. ╤ фЁєующ ёЄюЁюэ√, шьххЄё  Ёхчєы№ЄрЄ └.╚. ┴юсхэъю \cite{Bobenko} ш └.┬. ┴юыёшэютр \cite{Bolsinov}
юс ¤ътштрыхэЄэюёЄш ёыєўр  ╩ыхс°р-╧хЁхыюьютр эр $e(n)$ ёыєўр■ ╪юЄЄъш-╠рэръютр эр $so(n+1)$ (ёь. Єръцх \cite{BMMethods}).
╚ч ъюьсшэрЎшш ¤Єшї фтєї ЇръЄют ёыхфєхЄ, ўЄю чрфрўр ю ухюфхчшўхёъшї эр $(n-1)$-ьхЁэюь ¤ыышяёюшфх ¤ътштрыхэЄэр
ёыєўр■ ╪юЄЄъш-╠рэръютр эр $so(n+1)$.
┬ эр°хщ цх ъюэёЄЁєъЎшш ь√ яюыєўрхь {\it рёшьяЄюЄшўхёъє■} ёт ч№ чрфрўш ю ухюфхчшўхёъшї эр $(n-1)$-ьхЁэюь ¤ыышяёюшфх, сышчъюь ъ ёЇхЁх,
ё ёшёЄхьющ ╪юЄЄъш-╠рэръютр {\it ьхэ№°хщ ЁрчьхЁэюёЄш} -- эр рыухсЁх $so(n)$.

\section{╟ръы■ўхэшх}

╨рёёьюЄЁхэшх ухюфхчшўхёъшї эр ёырсю фхЇюЁьшЁютрээ√ї ёЇхЁрї тюёїюфшЄ  ъ ЁрсюЄх ╧єрэърЁх \cite{Poin}.
┬ эхщ яЁшьхэ хЄё  ъырёёшўхёъшщ ьхЄюф юёЁхфэхэш  ЄхюЁшш тючьє∙хэшщ ш т√тюф Єё  єёыютш , яЁш ъюЄюЁ√ї фрээр  ухюфхчшўхёър   ты хЄё  чрьъэєЄющ. ┬ фрээющ ЁрсюЄх ь√ ЁрёёьрЄЁштрхь тё■ ёютюъєяэюёЄ№ ухюфхчшўхёъшї
ш яюыєўрхь шї юяшёрэшх яюёЁхфёЄтюь рёшьяЄюЄшўхёъющ урьшы№Єюэютющ ЁхфєъЎшш шёїюфэющ ёшёЄхь√
ъ ёшёЄхьх ё ьхэ№°шь ўшёыюь эхшчтхёЄэ√ї.

╤ыхфє■∙хх трцэюх эрсы■фхэшх ърёрхЄё  ёт чш ¤Єшї тюяЁюёют ё шэЄхуЁры№эющ ухюьхЄЁшхщ.
╤ююЄтхЄёЄтє■∙шх шэЄхуЁры№э√х яЁхюсЁрчютрэш  с√ыш юЄъЁ√Є√ ╠шэъютёъшь, ╘єэъюь, ╨рфюэюь, ╔юэюь
ш т фры№эхщ°хь шёёыхфютрэ√ ш юсюс∙хэ√ эр ьэюуюьхЁэ√х чрфрўш ш фЁєушх ёшЄєрЎшш. 
╤є∙хёЄтхээ√щ тъырф т ¤Єє юсырёЄ№ ёюёЄрты ■Є ЄЁєф√ ╚.╠. ├хы№Їрэфр, \cite{Gelf}.
╤т ч№ яЁюсыхь√ ухюфхчшўхёъшї ё шэЄхуЁры№эющ ухюьхЄЁшхщ юЄъЁ√трхЄ тючьюцэюёЄ№ яЁшьхэхэш  ёююЄтхЄёЄтє■∙шї
Ёхчєы№ЄрЄют ш ьхЄюфют ъ фрээющ чрфрўх.
╬ЄьхЄшь, ўЄю ьхЄюф√ шэЄхуЁры№эющ ухюьхЄЁшш с√ыш яЁшьхэхэ√ ъ урьшы№Єюэют√ь ёшёЄхьрь ▌.┴. ┬шэсхЁуюь
т ЁрсюЄх \cite{Vinb}.

═ръюэхЎ, шьххЄё  яЁхфёЄрты ■∙шщ чэрўшЄхы№э√щ шэЄхЁхё ъырёё ёыєўрхт, фы  ъюЄюЁ√ї яюёЄЁюхээр 
ЁхфєъЎш  яючтюы хЄ яюыэюёЄ№■ юяшёрЄ№ Ёрёяюыюцхэшх ухюфхчшўхёъшї, шёяюы№чє  Єюяюыюушўхёъшх ьхЄюф√.
─ы  ¤Єюую рэрышчшЁєхЄё  Єюяюыюуш  ЄЁрхъЄюЁшщ ЁхфєЎшЁютрээющ ёшёЄхь√ (\ref{momHamEq}).
┬ ўрёЄэюёЄш, фы  фтєьхЁэ√ї фхЇюЁьшЁютрээ√ї ёЇхЁ Їрчютюх яЁюёЄЁрэёЄтю ЁхфєЎшЁютрээ√ї
ёшёЄхь -- фтєьхЁэр  ёЇхЁр, ш Єюяюыюуш  ЄЁрхъЄюЁшщ юяшё√трхЄё  Їрчют√ь яюЁЄЁхЄюь эр ¤Єющ ёЇхЁх.
┬ ёыєўрх ЄЁхїьхЁэ√ї ёЇхЁ ё юёхтющ ёшььхЄЁшхщ ЁхфєЎшЁютрээр  ёшёЄхьр хёЄ№ шэЄхуЁшЁєхьр 
ёшёЄхьр ё фтєь  ёЄхяхэ ьш ётюсюф√. ╥юяюыюуш  хх Ёх°хэшщ ьюцхЄ с√Є№ яЁюрэрышчшЁютрэр ё яюью∙№■ ьхЄюфют 
Єюяюыюушўхёъющ ъырёёшЇшърЎшш шэЄхуЁшЁєхь√ї ёшёЄхь.

└тЄюЁ сыруюфрЁшЄ ┬.╦. ├юыю чр тэшьрэшх ъ ¤Єющ ЁрсюЄх.

╨рсюЄр яюффхЁцрэр уЁрэЄрьш ╨╘╘╚  10-01-00748a, 11-02-01462a,
╧ЁюуЁрььющ яюффхЁцъш тхфє∙шї эрєўэ√ї °ъюы ═╪-3224.2010.1, ╧ЁюуЁрььющ ЁрчтшЄш  эрєўэюую яюЄхэЎшрыр,
уюёъюэЄЁръЄ 2.1.1.3704, ╘хфхЁры№эющ Ўхыхтющ яЁюуЁрььющ <<═рєўэ√х ш яхфруюушўхёъшх ърфЁ√ шээютрЎшюээющ ╨юёёшш>>
эр 2009-2013 уу., уюёъюэЄЁръЄ√ 14.740.11.0794, 02.740.11.5213.


\begin{thebibliography}{99}

\bibitem{jacoby}    C.G.J. Jacoby,
                    Vorlesungen \"uber Dynamik,
                    Reiner, Berlin (1884).

\bibitem{poinc}     H. Poincar\'e,
                    Les methodes nouvelles de la m\'ecanique c\'eleste,
                    Gauthier-Villars, Paris (1899).

\bibitem{DNF}       B. Dubrovin, S. Novikov, A. Fomenko,
                    Modern Geometry. Methods and Applications.
                    Springer-Verlag, Part 1, 1984; Part 2, 1985; Part 3, 1990.

\bibitem{Poin}      H. Poincar\'e,
                    Sur les lignes geodesique des surfaces convexes,
                    Trans. Amer. Math. Soc. 6 (1905), 237-274.

\bibitem{whittaker} E.T. Whittaker,
                    A Treatise on the Analytical Dynamics of Particles and Rigid Bodies, Ch. XIII,
                    Cambridge University Press, Cambridge (1917).


\bibitem{pohlmeyer} K. Pohlmeyer,
                    Comm.Math.Phys. {\bf 46}, 207 (1976).

\bibitem{Leggett}   J.A. Leggett,
                    Rev. Mod. Phys. 47,331 (1975).

\bibitem{golo}      V.L. Golo,
                    Lett. Math, Phys. {\bf 5},  155 (1981).


\bibitem{Kozlov}      V.V. Kozlov, Two integrable problems of classical mechanics,
                    Vestnik MGU, Ser. math. mech., 1981, N4, P. 80-83.

\bibitem{BMpaper}      A.V. Borisov, I.S. Mamaev,
                    Nonlinear Poisson brackets and isomorphisms in dynamics,
                    Reg. Chaot. Dynam., V 2, N 3/4, 1997, P. 72-89.

\bibitem{BMMethods}      A.V. Borisov, I.S. Mamaev,
                    Modern Methods of the Theory of Integrable Systems,
                    Moscow - Izhevsk: Institute of Computer Science, 2003. P. 296.

\bibitem{Perelomov}      A.M. Perelomov,
                    A note on geodesics on ellipsoid,
                    Reg. Chaot. Dynam., V 5, N 1, 2000, P. 89-94.


\bibitem{Schottky}     F. Schottky,
                    \"Uber das analytische Problem der Rotation eines starren K\"orpers in Raume von vier Dimensionen,
                    Sitzungsber. Konig. Preuss. Akad. Wiss. Berlin, 1891, Bd. XIII, S. 227-232. 

\bibitem{Manakov}     S.V. Manakov,
                    Note on the integration of Euler's equations of the dynamics of an n-dimensional rigid body,
                    Funct. Anal. Appl., 10(4), 328-329 (1976).


\bibitem{Bobenko}    A.I. Bobenko, 
                    Euler equations on the algebras e(3) and so(4). Isomorphism of the integrable cases,
                    Funktsional. Anal. i Prilozhen., 20 (1986), 64Ц66.

\bibitem{Bolsinov}    A.V. Bolsinov, 
                    Compatible Poisson brackets on Lie algebras and the completeness of families of functions in involution,
                    Math. USSR-Izv., 38:1 (1992), 69-90.

\bibitem{GS}        V. Golo, D. Sinitsyn,
                    Physics of Particles and Nuclei Letters, Vol. 5, No. 3, pp. 278-281 (2008).

\bibitem{klein}     F. Klein,
                    Vorlesungen \"uber h\"ohere Geometrie,
                    Berlin, Springer (1926).

\bibitem{Gelf}      I. Gelfand, S. Gindikin, M. Graev,
                    Selected topics in integral geometry
                    (Translations of Mathematical Monographs),
                    American Mathematical Society (2003).

\bibitem{HelgRad} S. Helgason,
                    The Radon Transform,
                    Second edition, Birkhauser Boston, 1999.

\bibitem{Vinb} E. B. Vinberg,
                      Equivariant symplectic geometry of cotangent bundles,
                      Mosc. Math. J., 2001, V 1, N 2, pp. 287-299.

\end{thebibliography}
\end{document}